# Electron transitions and local properties of $Nd^{3+}$ ion in $Nd_{0.5}Gd_{0.5}Fe_3(BO_3)_4$ single crystal

A.V. Malakhovskii[1,*], S.L. Gnatchenko[2], I.S. Kachur[2], V.G. Piryatinskaya[2], A.L. Sukhachev[1], V.L. Temerov[1]

[1] L. V. Kirensky Institute of Physics, Siberian Branch of Russian Academy of Sciences,
660036 Krasnoyarsk, Russian Federation

[2] B. Verkin Institute for Low Temperature Physics and Engineering,
National Academy of Sciences of Ukraine, 61103 Kharkov, Ukraine

*E-mail address: malakha@iph.krasn.ru

**Abstract**

Polarized absorption spectra and magnetic circular dichroism (MCD) spectra of $Nd_{0.5}Gd_{0.5}Fe_3(BO_3)_4$ single crystal were measured in the range of 10000 – 21000 $cm^{-1}$. The absorption spectra were studied as a function of temperature in the range of 2 – 293 K and as a function of magnetic field 0 – 65 kOe at 2 K. The excited $4f$ states of the $Nd^{3+}$ ion were identified in terms of the irreducible representations and in terms of $|J,\pm M_J\rangle$ wave functions of the free atom. The changes of the Landé factor during $f$-$f$ transitions and the exchange splitting of the ground and excited states were determined. It was found out that the selection rules for electron transitions substantially changed in the magnetically ordered state of the crystal. They are different for the different transitions and they strongly depend on the orientation of the Fe sublattice magnetic moment relative to the light polarization. It was revealed that the local magnetic properties in the vicinity of the excited $Nd^{3+}$ ion substantially depended on its state. In particular, in some excited states a weak ferromagnetic moment appears, the field induced reorientation magnetic transitions occur and the changes of the local magnetic anisotropy type take place. In some excited states the nonequivalent $Nd^{3+}$ centers were found out.

## 1. Introduction

The family of the rare-earth (RE) borates with the common formula $RM_3(BO_3)_4$, where M = Al, Ga, Fe, Cr, Sc and R – RE element, attracts considerable interest both in the fundamental aspect and in view of their manifold potential applications. In particular, the alumoborates possess the very good luminescent and nonlinear optical properties and can be used in the mini-lasers and in the lasers with the self-doubling frequency [1-3]. The growing interest to the RE ferroborates $RFe_3(BO_3)_4$ during the last years is stimulated by discovering of the multiferroic properties (i.e. correlation between magnetic, elastic and electric ordering) in many of them [4-8]. The multiferroic effects open the possibility of these materials usage in new multifunctional devices with the mutual control of magnetic, electric and elastic characteristics. From the viewpoint of the fundamental magnetism, RE ferroborates are of interest due to a wide variety of magnetic properties and phase transitions, which result from the presence of two interacting magnetic subsystems: iron and RE ones [9].

At high temperatures the RE ferroborates crystallize in the trigonal huntite-like structure with the space group $R32$ ($D_3^7$) [10-12]. The unit cell contains three formula units. The RE ions are located at the centers of the trigonal prisms $RO_6$ (the $D_3$ symmetry positions). The $Fe^{3+}$ ions occupy the $C_3$ positions in the octahedral environment of oxygen ions; these octahedrons form helicoidal chains along the $C_3$ axis. With the lowering temperature, some ferroborates with small ionic radius of RE ions undergo a structural phase transition to the $P3_121$ ($D_3^4$) symmetry phase [12]. It results in reducing of the RE ion position symmetry to the $C_2$ one and in appearance of two nonequivalent positions of $Fe^{3+}$ ions ($C_2$ and $C_1$). All RE ferroborates possess antiferromagnetic (AFM) ordering with the Neel temperature in the range of 30 – 40 K. The AFM ordering is conditioned by the exchange interaction within the iron subsystem [9]. The magnetic ordering of the RE subsystem is induced by the $f$-$d$ exchange interaction with the Fe-subsystem. The RE ions, in turn, due to their magnetic anisotropy, usually determine the direction of the $Fe^{3+}$ magnetic moments in the magnetically ordered state. The RE ferroborates can be easy-axis or easy-plane antiferromagnets. In some of them the reorientation phase transitions occur with the temperature change [13, 14].

The studied crystal has two nearest relatives: $GdFe_3(BO_3)_4$ and $NdFe_3(BO_3)_4$. The $GdFe_3(BO_3)_4$ borate demonstrates a rich phase diagram with a series of structural and magnetic phase transitions. At the temperature 156 K it undergoes a weak first-order structural phase transition with the lowering of the symmetry from the $R32$ to $P3_121$ space group [15, 12]. At $T_N$ = 38 K the AFM ordering of the crystal occurs, magnetic moments of $Fe^{3+}$ and $Gd^{3+}$ being aligned in the basal plane. Then, at the lowering temperature, the spontaneous spin-reorientation phase transition (of the first order) from the easy-plane to the easy-axis state takes place at 9 K [15]. For the $GdFe_3(BO_3)_4$ borate, the strong correlation between three order parameters (magnetic, electric and elastic) was established [4]. This allowed ones to consider this crystal as multiferroics.

The $NdFe_3(BO_3)_4$ crystal also demonstrates multiferroic properties [5, 8]. The $R32$ structure of this crystal remains down to the temperature of 1.6 K. Below $T_N = 30$ K the crystal orders antiferromagnetically with the orientation of $Fe^{3+}$ and $Nd^{3+}$ magnetic moments in the basal plane along the $C_2$ axes. So, three kinds of the equivalent magnetic domains can exist [16-18]. Below the temperature 13.5 K the commensurate magnetic structure transforms into a long-period antiferromagnetic helix with the single chirality [19-20]. This phase transition behaves as the first order one [21]. Optical spectra of the $Nd^{3+}$ ion in the $NdFe_3(BO_3)_4$ crystal in a wide spectral range (1500 – 25000 $cm^{-1}$) were studied in Ref. [22] and the crystal-field parameters and $g$-factors of the $Nd^{3+}$ states were obtained.

Magnetic phase transitions in the mixed borates $Nd_xGd_{1-x}Fe_3(BO_3)_4$ were studied by the optical spectroscopy method in Ref. [23]. It was found out that both the Neel temperature and the spin-reorientation transition temperature increase with the growth of the Nd concentration. The $Nd_{0.5}Gd_{0.5}Fe_3(BO_3)_4$ crystal, the same as the pure Nd and Gd-ferroborates, reveals multiferroic properties [24]. Magnetic properties of the crystal $Nd_{0.5}Gd_{0.5}Fe_3(BO_3)_4$ were studied in Ref. [25]. It was shown that below the $T_N = 32$ K it has an easy-plane AFM structure and the spin-reorientation does not occur down to 2 K. Any structural phase transitions were also not found down to 2 K. In the external magnetic field applied in the basal plane of the crystal a hysteresis in magnetization was found at 1 – 3.5 kOe. The hysteresis was observed at temperatures $T < 11$ K. Additionally, the temperature dependence of the magnetic susceptibility had a singularity at $T = 11$ K (when $H < 1$ kOe). These features were ascribed to the appearance of the static magnetic domains at $T < 11$ K. Spectroscopic studies in the external magnetic field $H \perp C_3$ at 2 K [26] also revealed some features indicating the presence of the domains.

The first measurements of the optical absorption spectra of the $Nd_{0.5}Gd_{0.5}Fe_3(BO_3)_4$ single crystal at temperatures 90 – 300 K as well as their analysis with the help of the Judd-Ofelt theory were performed in Ref. [27]. Some results of the study of the optical and magneto-optical properties of the $Nd_{0.5}Gd_{0.5}Fe_3(BO_3)_4$ crystal in the magnetically ordered state were presented in Refs. [26, 28]. A number of peculiarities of the selection rules for electron transitions in the ordered state of the crystal as well as the specific features of the local environment of the excited $Nd^{3+}$ ions were found.

In the present work the $f$-$f$ transitions from 11000 $cm^{-1}$ till the edge of the strong absorption at ~22000 $cm^{-1}$ are investigated. The crystal field split components of the $4f$ states are identified on the basis of absorption and magnetic circular dichroism (MCD) spectra. The MCD spectra allow finding not only the symmetry of the states but also their origin from the states of $|J, \pm M_J\rangle$ type of the free ion.

One of the main goals of this work is investigation of the influence of the *f-f* transitions on the local properties of the crystal. The local properties of compounds near the optically excited atoms (see, e.g., Refs. [29-33]) are of the growing interest in the recent years due to the problem of the optical computer creation (see e. g., Refs. [34-38]).

## 2. Experimental details

$Nd_{0.5}Gd_{0.5}Fe_3(BO_3)_4$ single crystals were grown from the melt solution on the base of $K_2Mo_3O_{10}$ as described in Ref. [39]. The absorption spectra were measured with the light propagating normal to the $C_3$ axis of the crystal for the light electric vector $\vec{E}$ parallel (the $\pi$ spectrum) and perpendicular (the $\sigma$ spectrum) to the $C_3$ axis and the light propagating along the $C_3$ axis (the $\alpha$ spectrum). The spectral resolution at the low temperature measurements was approximately equal to 1.5 cm$^{-1}$. The absorption spectra measured in the $\sigma$ and $\alpha$ polarizations coincide with each other within the limit of the experimental error. This implies that the absorption mainly occurs through the electric dipole mechanism.

Magnetic field was created by a superconducting solenoid with the Helmholtz type coils. The superconducting solenoid with the sample was placed in the liquid helium and all measurements in the magnetic field were fulfilled at $T = 2$ K. For the temperature measurements of absorption spectra a liquid-helium cooled cryostat was used. It had an internal volume filled by the gaseous helium where the sample was placed.

The MCD was measured in the field 5 kOe by the light-polarization modulation method using a piezoelectric modulator (details are in Ref. [40].) The spectral resolution of the MCD measurements was ~10 cm$^{-1}$ and sensitivity was $10^{-4}$. The MCD measurements were carried out in α-polarization.

## 3. Results and discussion

Absorption spectra of the $Nd_{0.5}Gd_{0.5}Fe_3(BO_3)_4$ crystal consist of the narrow bands corresponding to *f-f* transitions in $Nd^{3+}$ ions and of the wide bands due to *d-d* transitions in $Fe^{3+}$ ions [28]. The *d-d* spectra were subtracted from the total spectra and so the *f-f* spectra were obtained (Fig. 1). From the high energy side the studied spectra are restricted by the strong absorption conditioned by the Fe ions [28]. The *f-f* bands were identified according to Ref. [41].

Symbols of absorption bands (Fig. 1) were given according to Ref. [42]. Absorption spectra of the *f-f* transitions were studied in the temperature range of 2 – 300 K and in the magnetic field 0-65 kOe at the temperature 2 K. The MCD conditioned by the *d-d* transitions was not observed, and all the measured MCD spectra are due to the *f-f* transitions. The detailed absorption and MCD spectra of the *f-f* transitions will be given below.

### 3.1. Characterization of electronic transitions and states

#### 3.1.1. Background for characterization of electronic transitions and states.

Electron states in crystals are characterized by the irreducible representations in the group of the local symmetry ($D_3$ in our case). These characteristics of the states are found from polarization of transitions and from selection rules of Table 1. In crystals of the axial symmetry the electron states have one more characteristic: the crystal quantum number $\mu$. In trigonal crystals it has values [43]: $\mu$ = +1/2, -1/2, 3/2 (±3/2). Additionally, in the axial crystals the electron states can be described in a first approximation by $|J,\pm M_J\rangle$ wave functions of the free atom. Between values of $\mu$ and $M_J$ there is the following correspondence [43]:

$$M_J = \pm 1/2,\ \pm 3/2,\ \pm 5/2,\ \pm 7/2,\ \pm 9/2,\ \pm 11/2,\ \pm 13/2$$
$$\mu = \pm 1/2,\ (\pm 3/2),\ \mp 1/2,\ \pm 1/2,\ (\pm 3/2),\ \mp 1/2,\ \pm 1/2 \quad (1)$$

The states with $\mu = \pm 1/2$ correspond to the $E_{1/2}$ states and the states with $\mu = (\pm 3/2)$ correspond to the $E_{3/2}$ states in the $D_3$ group notations. Selection rules for the number $\mu$ in crystals are similar to those for the number $M_J$ in free atoms [43]. For the electric dipole absorption

$\Delta\mu = \pm 1$ corresponds to $\mp$ circularly polarized and $\sigma$-polarized waves,

$\Delta\mu = 0$ corresponds to $\pi$-polarized waves. (2)

For the linearly polarized waves, these selection rules coincide, of course, with that of Table 1.

According to the definition, the splitting of the Kramers doublets in the magnetic field directed along the $C_3$ axis of a crystal is:

$$\Delta E = \mu_B g_C H. \quad (3)$$

Here $g_C$ is the effective Landé factor in the $C_3$-direction. The same value in the approximation of the $|J,\pm M_J\rangle$ function is evidently defined by the equation:

$$\Delta E = 2g\mu_B M_J H, \quad (4)$$

where $g$ is the Landé factor of the free atom. Correspondingly, the Landé factor of the Kramers doublet along the $C_3$ axis in the same approximation is

$$g_{CM} = 2gM_J \quad (5)$$

These values for the states of $Nd^{3+}$ ion are given in Table 2. The states with the same $\mu$ and different $M_J$ (see Eq. 1) can mix in the crystal, and the resulting $g_C$ can differ from $g_{CM}$. The prevailing $M_J$ state of the free atom in the crystal field state can be found in a first approximation (Table 3) basing on the comparison of $g_{CM}$ for the corresponding $M_J$ (Table 2) with the theoretical $g_C$ in the $NdFe_3(BO_3)_4$ crystal (Table 3). The MCD spectra can help to identify states in the $|J,\pm M_J\rangle$ functions approximation and to find the Zeeman splitting of lines.

The MCD conditioned by a pair of the Zeeman splitting components is evidently described by the equation:

$$\Delta k = k_{m+}\varphi(\omega,\omega_0+\Delta\omega_0) - k_{m-}\varphi(\omega,\omega_0-\Delta\omega_0) \qquad (6)$$

Here $k_+$ and $k_-$ are the amplitudes of (+) and (-) circularly polarized lines, $\varphi$ are the form functions of (+) and (-) polarized lines. If the Zeeman splitting $\Delta\omega_0$ is much less than the line width, then one obtains

$$\Delta k = k_m c \varphi(\omega,\omega_0) + k_m \Delta\omega_0 \partial\varphi(\omega,\omega_0)/\partial\omega_0 . \qquad (7)$$

Here $k_m = k_{m+} + k_{m-}$ is amplitude of the line not split by the magnetic field and $c = (k_{m+} - k_{m-})/k_m$. The first term in (7) is the paramagnetic MCD and the second one is the diamagnetic MCD.

The fine structure of the MCD spectra is conditioned by the diamagnetic effect. The Zeeman splitting of an absorption line $2\Delta\omega_0$ is found through the Zeeman splitting of the initial and final levels (the Kramers doublets in particular):

$$2\hbar\Delta\omega_0 = \pm(\Delta E_i \pm \Delta E_f) = \mu_B H \Delta g_C . \qquad (8)$$

The sign of the temperature dependent paramagnetic effect ($c$ in eq. 7) is defined by polarization of the transition from the lower component of the Zeeman splitting of the initial state. Sign of the diamagnetic effect ($\Delta\omega_0$ in eqs. (7)) is not so unambiguous. The first sign in (8) is the sign of the paramagnetic effect and the second one indicates that the splitting of the line can be sum or difference of the splitting of the initial and final states (see below).

The MCD spectra allow us to find the Zeeman splitting of lines and, consequently, the change of the Landé factor during the electron transition. In the case of the Gaussian shape of the absorption line, the Zeeman splitting is found with the help of the formula [44]:

$$\Delta\omega_0 = \frac{\Delta k_{dm}}{k_m}|\omega_m - \omega_0|\sqrt{e} . \qquad (9)$$

Here values $\Delta k_{dm}$ and $\omega_m$ are amplitude and position of extremum of the diamagnetic MCD, respectively, $k_m$ is amplitude of the $\alpha$-polarized absorption line. For the Lorentzian form function $\sqrt{e}$ should be replaced by "2". Formula (9) can, evidently, be used only for the solitary, well resolved pair of the Zeeman splitting components. The splitting $\Delta\omega_0$ and the corresponding values $\Delta g_C$ according (8) were found from the experimental spectra of the absorption and MCD, where it was possible (see Table 3).

The general schematic diagram of the 4*f* states and *f-f* transitions and their circular polarizations in the magnetic field directed along the $C_3$ axis (see Fig. 2) can be created with the help of (1) and (2), taking into account in a first approximation the splitting (4) in the magnetic field according to $\pm M_J$ and the splitting in the CF according to $|M_J|$. The real positions of states with the different values of the $M_J$ will be found below on the basis of the experimental data analysis. When the splitting of an absorption line in the magnetic field is equal to the difference of the splitting of the states (sign minus in (8) in brackets) and these splitting are close in value, the supposed sign of the line splitting is not definite.

### 3.1.2. $^4I_{9/2}\rightarrow{}^4F_{3/2}$ transition (R band)

Preliminary results concerning the R band were presented in Ref. [28]. The polarized absorption and MCD spectra of the transition $^4I_{9/2}\rightarrow{}^4F_{3/2}$ are depicted in Fig. 3. At the liquid helium temperature only lines R1 and R2 are observed in the R band. According to the polarization of these lines and selection rules of Table 1, the symmetry of the lowest level of the ground multiplet (Gr1) is $E_{1/2}$ and symmetries of the levels R1 and R2 are $E_{1/2}$ and $E_{3/2}$, respectively (Table 3). The R(2-1) and R(3-1) lines (Fig. 3) correspond to the transitions from the levels Gr2 and Gr3 at the energies 79 and 150 cm$^{-1}$, respectively, to the R1 state (Table 3). From the polarization of the R(2-1) and R(3-1) lines and from the selection rules of Table 1 it follows that the Gr2 and Gr3 levels have symmetries $E_{3/2}$ and $E_{1/2}$, respectively (Table 3). The R(2-1) line can be identified also as the transition Gr3→R2 according to its position and the selection rules. However, the temperature dependencies of the R(2-1) and R(3-1) lines intensities testify that the former assumption is more correct.

Signs of the diamagnetic MCD (signs of $\Delta\omega_0$) are easily found from Fig. 3 according to the definition (6) (they are shown in Fig. 3 and Table 3). There are at least two suitable candidates for the ground state with $E_{1/2}$ symmetry: $M_J=\pm 7/2$ ($\mu=\pm 1/2$) and $M_J=\pm 5/2$ ($\mu=\mp 1/2$) (Fig. 2). However, according to Fig. 2, only in the case of the ground state $M_J=\pm 5/2$ ($\mu=\mp 1/2$) the diamagnetic MCD of lines R1 and R2 has experimentally observed signs (Fig. 3). Despite the trigonal symmetry of the Nd$_{0.5}$Gd$_{0.5}$Fe$_3$(BO$_3$)$_4$ crystal, it has the easy plane magnetic anisotropy in the magnetically ordered state [25], i. e., the crystallographic single axial anisotropy is not strong. As a consequence, the order of the levels and their Landé factors can not correspond to the value of the magnetic quantum number $M_J$ of the free atom. The Gr3 level (Table 3, Fig. 2) can be referred to the state $M_J=\pm 7/2$ ($\mu=\pm 1/2$). Indeed, according to Fig. 2, the transitions from the level $M_J=7/2$ should have opposite circular polarizations relative to those of the transitions from the lowest level $M_J$=5/2, and the line R(3-1) (Gr3→R1 transition) should have negative diamagnetic MCD ($\Delta\omega_0$<0), that is really observed (Fig. 3). The R(2-1) line (Gr2→R1 transition) should have positive diamagnetic effect. This absorption line has asymmetric shape (Fig. 3). The most probably, it is due to the vibronic sideband which can have the diamagnetic MCD of the opposite sign relative to that of the electronic origin. This can be the reason of the small MCD of the R(2-1) line (Fig. 3). Some lines in the *D* band have the similar shape (see below).

With the help of the procedure described above, the Zeeman splitting of the R2 line (in the field *H*=1 kOe) was found: $2\Delta\omega_0$=-0.16 cm$^{-1}$ and the corresponding change of the Landé factor is: $\Delta g_C$=-3.4 (Table 3). Since $\Delta g_C$ is proportional to $\Delta\omega_0$ we ascribe sign of $\Delta\omega_0$ to $\Delta g_C$. According to (8), diagram of Fig. 2 and Table 2 the change of the Landé factor in the $|J,\pm M_J\rangle$ function approximation is: $\Delta g_{CM}$= -(1.2+3.64)= -4.84 (Table 3). Sum of the absolute values of the theoretical $g_C$ in NdFe$_3$(BO$_3$)$_4$ (Table 3) for this transition is 2.938.

### 3.1.3. $^4I_{9/2}\rightarrow{}^2H_{9/2}+^4F_{5/2}$ transitions (S band)

Excited state of the S transition is split in the following way: $^4F_{5/2}$: $2E_{1/2}+E_{3/2}$ and $^2H_{9/2}$: $3E_{1/2}+2E_{3/2}$. Symmetries of states in the S manifold (Table 3) are found according to the linear polarizations of the absorption lines (Fig. 4), selection rules of Table 1 and above identification of the ground state. Signs of the Zeeman splitting $\Delta\omega_0$ of lines in the S band are not so much evident from the MCD spectra (Fig. 5) as in the R band. It is possible to show [44] that signs of extremums of the $\partial\Delta k/\partial\omega$ function at the absorption line positions give signs of the diamagnetic effect (see Fig. 5 and Table 3). The prevailing $M_J$ states of the free atom in the crystal field states of the S manifold (Table 3) can be preliminary found basing on the comparison of the Landé factors $g_{CM}$ for the corresponding $M_J$ with the theoretical $g_C$ in the NdFe$_3$(BO$_3$)$_4$ crystal (Table 3 and 2). They are of course different but the succession of the values permits to identify the origin of the S states from the $M_J$ states.

The genetic origination of the crystal field states from the $M_J$ states of the free atom on the ground of the diamagnetic MCD signs is not unambiguous for this manifold. According to the diagram of Fig. 2, the transitions $E_{1/2}$ (*M*=5/2)→$E_{3/2}$ (*S*3, *S*6 and *S*8 lines) can have only negative signs of the diamagnetic MCD but in the S8 line it is positive (Fig. 5, Table 3). Additionally, the lines S(3-1) and S1 should have opposite signs of the diamagnetic effect, the same as it takes place for R(3-1) and R1 lines (see above). However, the lines S(3-1) and S1 have diamagnetic effects of the same sign (Fig. 5). The Zeeman splitting $2\Delta\omega_0$ of lines S2, S4, S7 and S8 were found as described above. They are: +0.39, -0.154, +0.23 and +0.647 cm$^{-1}$, respectively. Corresponding changes of the Landé factors are given in Table 3. The theoretical changes of the Landé factors in the $|J,\pm M_J\rangle$ function approximation were calculated according to the identification of the $M_J$ value of the excited states (Table 3), diagram of Fig. 2 and eq. (8). Comparison of the theoretical $\Delta g_{CM}$ with the measured ones permitted us to refine identification of the S4 and S5 states (see Table 3). The theoretical value for S8 line is close in absolute value to the experimental one but have the opposite sign.

### 3.1.4. $^4I_{9/2} \rightarrow {}^4S_{3/2} + {}^4F_{7/2}$ transitions (A band)

Absorption spectra of the A band at 2 K are shown in Fig. 6. The excited state is split in the following way: $^4S_{3/2}$: $E_{1/2}+E_{3/2}$ and $^4F_{7/2}$: $3E_{1/2}+E_{3/2}$. Quantity of the observed lines corresponds to this splitting. Assignment of A1, A2, A5 and A6 states based on the polarizations of the corresponding lines (Fig. 6) is unambiguous. Polarization of the A3 and A4 lines is not evident, but there is no other possibility for their assignment as shown in Table 3. Origination of the A states from the $|J,\pm M_J\rangle$ states was supposed, comparing the theoretical values of $g_C$ in NdFe$_3$(BO$_3$)$_4$ with $g_{CM}$ (Table 3). Large difference between these values for the A3 and A4 states is surprising. Probably it is connected with the not strong axial anisotropy and, as a consequence, with the large value of $g_\perp$ (Table 3). The A1 and A5 lines are very weak in the $\sigma$-polarization at 90 K (Fig. 7), therefore they give very small contribution into the MCD, and practically all MCD is due to A2 and A6 lines (Fig. 7). The Zeeman splitting and corresponding $\Delta g_C$ for these transitions were found as described above (see Table 3). Values of $\Delta g_{CM}$ were also calculated (Table 3). Observed $\Delta g_C$ are close to $\Delta g_{CM}$ and their signs correspond to the identification according to the diagram of Fig. 2. The assignment of transitions from the upper sublevels of the ground state (Fig. 7) corresponds to their linear polarizations.

### 3.1.5. $^4I_{9/2} \rightarrow {}^4G_{5/2} + {}^2G_{7/2}$ transitions (D band)

Excited state of the D manifold is split in the following way: $^4G_{5/2}$: $2E_{1/2}+E_{3/2}$ and $^2G_{7/2}$: $3E_{1/2}+E_{3/2}$. Symmetries of states in the D manifold are found (Table 3) according to the linear polarizations of the absorption lines (Fig. 8, Table 3), selection rules of Table 1 and above identification of the ground state. Signs of the circular polarizations of the lines (Table 3) were found with the help of the first derivative of the MCD spectrum (Fig. 9). Unfortunately, it was possible to find the Zeeman splitting $\Delta\omega_0$ and corresponding $\Delta g_C=+5.6$ only for D(4-1) transition (Fig. 9) from the Gr4 sublevel of the ground state (Table 3). $\sigma$-polarization of this line permitted to identify symmetry of the Gr4 state as $E_{3/2}$ (see Table 3). According to the identification of the D(4-1) transition, $\Delta g_{CM}=-1.61$ for it. Shapes of D(2-1) and D(2-2) lines (Fig. 9) are similar to that of the R(2-1) line (Fig. 3) and probably they are also due to the vibronic side-bands.

### 3.1.6. $^4I_{9/2} \rightarrow {}^4G_{9/2}+{}^4G_{7/2}+{}^2K_{13/2}$ transitions (E+F-bands)

Excited state of the (E+F) manifold is split in the following way: $^4G_{9/2}$: $3E_{1/2}+2E_{3/2}$, $^4G_{7/2}$: $3E_{1/2}+E_{3/2}$ and $^2K_{13/2}$: $5E_{1/2}+2E_{3/2}$. Corresponding sets of the transitions are $3\pi\sigma+2\sigma$, $3\pi\sigma+\sigma$ and $5\pi\sigma+2\sigma$. According to the observed polarizations of the absorption lines (Fig. 10) and to the separate position of the E1-E5 lines it is logical to refer them to the $^4G_{9/2}$ multiplet (Table 3). It is difficult to separate two other multiplets in the F band (Fig. 10). Additionally, one $\sigma$-polarized line of the transition into the $E_{3/2}$ state is not observed in this band. Signs of $\Delta\omega_0$ ($\Delta g_C$) were found (where it was possible) from the $d\Delta k/dE$ function (Fig. 11). It was possible to find values of $\Delta\omega_0$ ($\Delta g_C$) only for the lines F1 and F6 (see Table 3). Necessary amplitudes of absorption lines were found as a result of decomposition of the absorption spectrum into the Lorentz components. The F1 state was identified as $M_J=11/2$ ($^2K_{13/2}$) state since only in this case the theoretical $\Delta g_{CM}$ is positive and is larger than the measured one.

All results obtained above require some comments. Experimentally observed linear polarizations of lines and symmetry of states are unambiguously connected via selection rules of Table 1. CF mixes states with the same symmetry and so it does not violate selection rules for the linear polarizations of transitions. The situation for the circular polarizations is more complicated. Axially symmetric CF mixes states with the same crystal quantum number μ. However, numbers $M_J$ of the different signs can correspond to the states with the same $\mu$ (see Eq. (1)). Consequently, transitions with the participation of such states will have opposite circular polarizations, since the splitting in the magnetic field occurs according to $M_J$, but the polarization of transitions is governed by the number $\mu$ (see eq. (2)). Majority of the experimentally found $\Delta\omega_0$ ($\Delta g_C$) correspond by sign and even by value to the supposed $|J,\pm M_J\rangle$ function (Table 3), but there are some transitions which contradict to it. This contradiction for the $E_{1/2}$ excited states can be due to the mixing of states with the same number $\mu=\pm 1/2$ and different values and signs of $M_J$, which give different signs of $\Delta\omega_0$. Additional contribution into the discrepancies can give the mixing of the close manifolds which present in the S, A, D, E and F bands. Such explanation is not suitable for the transitions Gr1 (J=±5/2, μ=∓1/2)→$E_{3/2}$, since all these transitions should have negative $\Delta\omega_0$ (Fig. 2). However, if we take into account, that the ground state Gr1 is also a mixture of states with the different $M_J$, then the same explanation of the discrepancy will be possible. In conclusion it is worth noting that the above consideration refers to the paramagnetic state of the crystal.

## 3.2. Behavior of absorption lines as a function of the magnetic field and temperature

### 3.2.1. Temperature behavior of transitions into the $E_{3/2}$ states

The characteristic feature of the $E_{3/2}$ states is that $g_\perp=0$ in these states (see Table 3). Only two of the corresponding absorption lines revealed a temperature dependent splitting below the Neel temperature (Fig. 12). This splitting should be equal to the splitting of the $Nd^{3+}$ ground state due to the exchange interaction with the iron ions, because magnetic moments of the iron sublattice are in the plane perpendicular to the $C_3$ axis and $g_\perp=0$ in the considered excited states of the $Nd^{3+}$. We supposed the exchange splitting of the ground state to be average of values 10 and 8.8 cm$^{-1}$ at 6 K presented in Fig. 12, i.e., 9.4 cm$^{-1}$, that corresponds to $H_{ex}$=84.4 kOe, taking into account the theoretical value $g_\perp$=2.385 (Table 3). The exchange splitting of the $Nd^{3+}$ ion ground state found in $NdFe_3(BO_3)_4$ was 8.8 cm$^{-1}$ [22].

The different values of the exchange splitting found from the different transitions mean that the electron transition has an influence on the local properties of the ground state as well. The similar situation we saw above, when we found positions of states of the ground multiplet from the different transitions (See Figs. 3, 5, 7, 9, 11. In Table 3 the average values are given). Indeed, electron transitions occur due to mixing of initial and final states by the time dependent perturbation caused by the electromagnetic wave. Similar phenomenon was earlier noted also in Ref. [44]. The singularities in the temperature dependences of Fig. 12 in the region of 14-16 K testify that there are some local distortions in the corresponding excited states. Not all transitions demonstrate the splitting connected with the exchange splitting of the ground state. This means that the transition from the upper sublevel of the ground state exchange splitting is forbidden.

### 3.2.2. $^4I_{9/2} \rightarrow {}^4F_{3/2}$ transition (R-band)

The geometry of measurements in the magnetic field is presented in Fig. 13. In the R1 state $g_\perp \neq 0$ (Table 3) and the exchange splitting is possible. Therefore, there are possible four transitions between components of the exchange splitting of the ground state and the excited R1 state (see Fig. 14). However, no splitting of the R1 line was observed both at 2 K and at temperatures between 2 K and the Neel temperature. The difference of the σ- and π-polarized lines energies seen at $H$=0 kOe (Fig. 15) is apparently connected with the absorption of the $Nd^{3+}$ ions positioned in domains and domain walls. At $H_\perp$=4 kOe the crystal is in one domain state [25] and this difference disappears. (The splitting in the large magnetic field (Fig. 15) will be discussed in the section **3.2.5.**) The lack of the R1 line splitting at $H_\perp$=4 kOe ($T$=2 K) and as a function of temperature is possible in several cases. 1) The exchange splitting of the excited and ground states are equal and R1c and R1d transitions are forbidden (Fig. 14). 2) The exchange splitting of the excited state is equal to zero and R1b and R1d transitions are forbidden. The lack of the exchange splitting of the excited state is hardly probable since $g_\perp \neq 0$ in this state (Table 3). 3) Only the R1a transition is allowed. The latter variant is preferable. $H$=4 kOe is also the field of the spin-flop, i. e., magnetic moments of ions become perpendicular to the magnetic field. Then Fig. 15 inset testifies that σ absorption is more probable when magnetic moments are perpendicular to the σ-polarization (Fig. 13). The π-polarization is always perpendicular to the magnetic moments.

The splitting of the R2 line both in $H \perp C_3$ and $H \| C_3$ was not observed. For $H \perp C_3$ it is natural since $g_\perp$=0 in the excited state (Table 3).

### 3.2.3. $^4I_{9/2} \rightarrow {}^2H_{9/2} + {}^4F_{5/2}$ transitions (S-band)

**S2 line**

This line can be studied only in π-polarization because of high intensity of the S2σ line (see Fig. 4). The experiments in magnetic field were fulfilled at $T$=2 K. At this temperature only transitions from the lowest level of the ground state, S2a and S2c, are observed (Figs. 16, 17, 18). Fig. 17 demonstrates transformation of the S2π transition spectrum in the field $H \| C_3$. In the field $H \perp C_3$ the energies of the S2aπ and S2cπ transitions change slightly and identically, but in the field $H \| C_3$ the splitting increases (Fig. 18). This behavior is apparently due to the strong magnetic anisotropy. In particular, as opposed to the ground state, the $C_3$ direction is the easy axis in the S2 state. If the magnetic field is applied along the easy direction, then the magnetic moments of the antiferromagnet orientate perpendicular to the magnetic field. In this case the splitting of the S2 doublet in magnetic field (Fig. 18, inset) should be described by the formula:

$$\Delta E = \mu_B g_M \left( H_{Fe}^2 + H^2 \right)^{1/2} \tag{10}$$

in assumption that orientation of the Fe magnetic moments remains purely antiferromagnetic. $H_{Fe}$ in (10) is the exchange field Fe-Nd and $g_M$ is the Landé factor of the $Nd^{3+}$ ion in the direction of its magnetic moment in the excited state. At the large magnetic field $H$ the formula (10) transforms into the straight line going through the origin of coordinates, however in reality it is not so (see Fig. 18, inset). The dependence of the Fig. 18, inset can be obtained if we substitute $H$ for $H+\Delta H$, where $\Delta H$=constant. This value $\Delta H$ corresponds to some exchange field and spontaneous magnetic moment along the external field.

A temperature dependent splitting of the S2π line is also observed [28]. At $T$=6 K it is 9.6 cm$^{-1}$ (Table 3). It is just between two values of the ground state exchange splitting mentioned above. Additionally, during decomposition of the S2 line spectrum as a function of temperature, it is impossible to differentiate S2aπ and S2cπ lines and S2bπ and S2dπ lines (Fig. 16) because of the small splitting of the excited state. Therefore, the diagram of Fig. 16 is not totally definite. In particular, final states of the S2 transition can be rearranged. Then, in contrast to Fig. 16, the splitting between the S2aπ and S2bπ lines will be equal to the sum of the exchange splitting of the ground and excited states. We can not choose the variant since the excited state exchange splitting is small and, as mentioned above, value of the found ground state exchange splitting depends on the transition from which it is found.

**S4-line**

Fig. 19 demonstrates transformation of the $\sigma$-polarized S4 line spectrum in the magnetic fields $H \perp C_3$ and $H \| C_3$. The behavior of the S4 lines positions in the field $H \perp C_3$ is shown in Fig. 20. In the one domain state of the crystal ($H_\perp > 4$ kOe) the energies of the S4a$\sigma$ and S4a$\pi$ lines are equal. Difference of these energies at $H_\perp < 4$ kOe, as mentioned above, can be connected with the absorption in domains and domain walls. At $H_\perp > 20$ kOe the $\sigma$-polarized line S4c$\sigma$ appears, but S4c$\pi$ line does not appear (Fig. 20). (See also diagram of Fig. 21.) The asymmetric changes of the S4a$\sigma$ and S4c$\sigma$ transitions energies as a function of the magnetic field $H \perp C_3$ (Fig. 20) is probably connected with the change of the ground state energy.

The behavior of the S4 lines positions in the field $H \| C_3$ is presented in Fig. 22. In this field the S4c line appears in both polarizations already at ~10 kOe. The magnetic field $H \| C_3$ does not create the one domain state. Therefore, a small splitting connected only with the domain structure is observed in the weak field (Fig. 22). It is rather surprising that in one domain state at the low magnetic field there is no exchange splitting of the excited state, since according to the theory $g_\perp \neq 0$ in this state (Table 3). This means that there is no Fe-Nd exchange interaction in the considered excited state of the Nd$^{3+}$ ion. The observed splitting 11.15 cm$^{-1}$ in the field $H_\| = 65$ kOe is actually the splitting of the paramagnetic ion. Corresponding $g_C = 3.68$ is of the same order of magnitude as the theoretical values of $g_C$ and $g_{CM}$ (Table 3). The splitting in the field H$_\perp$ corresponds to g$_\perp$=1.64 that is also close to the theoretical value in the NdFe$_3$(BO$_3$)$_4$ crystal (Table 3).

The temperature dependent splitting of the S4 line both in $\pi$ and $\sigma$ polarizations is also observed (Fig. 23). Their values at 6 K are 8 and 9 cm$^{-1}$, respectively (Table 3). They are a little different, since the experiment is made in the zero magnetic field, i. e., in the many domain state (see Fig. 20). They are also close to the found above exchange splitting of the ground state, because the real exchange splitting of the S4 excited state in the homogeneous state of the crystal is equal to zero.

**S5 and S6-lines**

Absorption spectra of the S5 and S6 lines in the magnetic field $H\perp C_3$ are shown in Fig. 24. Positions of S5 lines as a function of the magnetic field are presented in Fig. 25. The S5 line is not split at $H=0$ (Fig. 25), i.e., there is no exchange splitting in spite of $g_\perp\neq 0$ in the S5 state (Table 3), and the $Nd^{3+}$ ion is actually in the paramagnetic state. The observed splitting in the external field $H\perp C_3$ (Fig. 25) corresponds to $g_\perp=4.0$. This value is very close to the theoretical $g_\perp$ in $NdFe_3(BO_3)_4$ crystal (Table 3). The zero splitting of the S5 line in the magnetic field $H\|C_3$ (Fig. 25) testifies that in the S5 state there is a strong local one-ion magnetic anisotropy of the easy plane type. The S6 line is not split both in $H\perp C_3$ and $H\|C_3$. It is quite natural that the S6 line is not split in the exchange field and in the external field $H\perp C_3$ since theoretically $g_\perp=0$ in the S6 state (Table 3). However, in $H\|C_3$ the S6 line could split in the external field. The absence of the splitting testifies to the strong local crystallographic anisotropy in the S6 state. The temperature dependent splitting of the S5 and S6 lines is not observed.

**S7-line**

Transformation of the S7 line spectrum in the magnetic field $H\perp C_3$ is depicted in Fig. 26 inset. In the one domain state ($H=4$ kOe) the S7 line is not split by the exchange field. Consequently, $Nd^{3+}$ ion in the S7 excited state is in the paramagnetic state, in spite of the not zero theoretical $g_\perp$ in the $D_3$ symmetry (Table 3). The splitting in the magnetic field $H\perp C_3$ is observed only in the π-polarization and between π and σ-polarizations (Fig. 26), i.e., S7cσ transition is forbidden (notations of transitions is similar to the used above). The splitting in the field $H_\perp=65$ kOe corresponds to $g_\perp=3.47$ close to the theoretical value in $NdFe_3(BO_3)_4$ (Table 3). The non linear splitting as a function of $H_\perp$ (Fig. 26) is, probably, due to the anisotropy in the basal plane.

In the magnetic field $H\|C_3$ the splitting of the S7 line is not observed in spite of the theoretical $g_C\neq 0$ in the S7 state (Table 3) that testifies to the strong anisotropy. The temperature dependent splitting of the S7 line is also not observed.

**S8-line**

This line exists only in the σ-polarization (Fig. 4). Its energy behavior in the fields $H\perp C_3$ and $H\|C_3$ is presented in Fig. 27. The exchange splitting ($H=0$) is not observed in accordance with the theoretical $g_\perp=0$ in the S8 state (Table 3), and, consequently, the $Nd^{3+}$ ion is in the paramagnetic state. The splitting 24.3 cm$^{-1}$ in the field $H_\|=65$ kOe (Fig. 27) corresponds to $g_C=8.0$ that is close to the theoretical values of $g_C$ and $g_{CM}$ (Table 3). The behavior of the S8 line in the fields $H \perp C_3$ and $H \| C_3$ (Fig. 27) is opposite to that of the S5 line (Fig. 25). In particular, these dependences testify that the magnetic anisotropy of the $Nd^{3+}$ ion in the S8 state is of the easy axis type but in the S5 state it is of the easy plane type. The temperature dependent splitting of the S8σ line was shown in Fig. 12.

### 3.2.4. $^4I_{9/2} \rightarrow {}^4S_{3/2} + {}^4F_{7/2}$ transitions (A-band)

**A1 line**

The A band spectra were presented in Fig. 6. According to Table 3, $g_\perp \neq 0$ in the A1 state. Therefore, the splitting of the A1 state in the exchange field of the iron sublattice can exist. However, it is very small (Fig. 28). $Nd^{3+}$ ions moments in the A1 state till ~22 kOe (Fig. 28) are apparently perpendicular to the magnetic field $H \perp C_3$. The energies of the exchange split states change almost identically due to the field of anisotropy, which is parallel to the moments. Since the exchange field in the A1 state is very small, the not large field of anisotropy is enough for such effect.

Above 22 kOe the energies of the A1 doublet change in opposite directions. This magnetic state corresponds to orientation of the magnetic moments along the external magnetic field $H \perp C_3$ (Fig. 29). The splitting 10.95 cm$^{-1}$ at 65 kOe corresponds to $g_\perp=3.62$ if the field is counted from zero. If the field is counted from the point of the crossing of the lines in Fig. 28, than the splitting corresponds to $g_\perp=5.46$ (compare with the theoretical value 3.216 in Table 3). Polarizations of the transitions into the A1a and A1c states are different (see Figs. 28, 29) in contrast to what it should be according to symmetry of the A1 state in the $D_3$ local symmetry. The latter takes place, e. g., in S5 transition (Fig. 25). The A1 line does not split with the decreasing temperature below $T_N$. This means that transitions from the upper sublevel of the ground state exchange splitting are forbidden. The splitting of the A1 line in $H \| C_3$ is not observed that testifies to the strong magnetic anisotropy of the $Nd^{3+}$ ion in the A1 excited state.

**A5-line**

The A5π line (Fig. 6) is split with the decreasing temperature below $T_N$ (Fig. 30) due to the exchange splitting of the ground state. This splitting is 16.4 cm$^{-1}$ at 6 K, i. e., it is substantially larger than the exchange splitting of the ground state being equal to 9.4 cm$^{-1}$. This is possible with the energy diagram of Fig. 31 with the exchange splitting of the excited state of ~7 cm$^{-1}$. There are two variants of the transitions diagram from the view-point of the energetically favorable orientation of the sublattice magnetic moment in the excited state. In the first one it is the same as that in the ground state and then the transitions A5a and A5b occur with the overturn of the magnetic moment direction. In the second variant everything is vice-versa (Fig. 31). We have chosen the second variant. In this case, transitions A5c and A5d occur with the change of the magnetic moment direction. The A5 line is not split in the magnetic field $H\perp C_3$ (Fig. 32). Thus, the A5c line (diagram of Fig. 31) is not observed. There is also no indication of the A5d line observation. At $H$=25 kOe a kink on the field dependence of the A5a line position is observed similar to that observed on the field dependence of the A1 line position (Fig. 28). This can testify to the similar change of the local magnetic state in this excited electron state.

**A2 and A6 lines**

These lines, which are not active in π-polarization, appear in this polarization in the magnetic field $H\perp C_3$ (Fig. 32, inset). This means that the changed magnetic structure violates the axial symmetry. The A2 and A6 states have $g_\perp$=0 (Table 3). Therefore, the splitting of the corresponding absorption lines both in the exchange field and in the external magnetic field $H\perp C_3$ should not exist. Unfortunately, the A2σ and A6σ lines are too strong to be studied. However, a kink on the field dependence of the A6π line position again testifies to some change of the local magnetic state in this excited electron state.

In the field $H||C_3$ the π-polarized A band spectra practically are not changed that testifies to the strong easy plain anisotropy. The σ-polarized spectra can not be analyzed because of their high intensity.

### 3.2.5. $^4I_{9/2}\to{}^4G_{5/2}+{}^2G_{7/2}$ transitions (D-band)

This absorption band was studied in detail in the previous work [26]. Here we will mention only some results. The total spectrum of this transition was shown in Fig. 8. Intensities of some transitions between components of the ground and excited state exchange splitting strongly depend on the orientation of the magnetic moments. In particular, intensity of the D1cσ transition (see diagram of Fig. 33) becomes zero at $H_\perp$=4 kOe (Fig. 34). In this field the crystal proceeds to the one domain state with the magnetic moments perpendicular to the magnetic field [25]. According to the geometry of the experiment (Fig. 13), this means that the D1cσ transition is allowed only for the magnetic moments $M||\sigma$ (the D1cπ transition is not observed at all). At the increasing field such magnetic moments appear again and, therefore, the D1cσ line also appears (Fig. 34). Additionally, the D1cσ line reveals a peculiar behavior as a function of temperature (Fig. 35): it disappears at temperature $T$>20 K. This was explained by the appearance of a weak ferromagnetic moment in the excited state at $T$<20 K. The D1b line (diagram of Fig. 33) was observed only in the natural circular dichroism (NCD) spectrum [26].

The D5c$\pi$ line appears only in the magnetic field $H||C_3$ at $H_{||}$> 30 kOe (Fig. 36), i. e., it is allowed only when $M||\pi$. The field dependence of the D2 line splitting (Fig. 37, inset), testifies to appearance of a spontaneous magnetic moment along the magnetic field according to the logic discussed above.

There was an unexpected observation that the energies of some transitions depended on the polarization. In particular, in the region of 0-4 kOe this occurs in the D1 transition, and at $H$>4 kOe (one domain state) this occurs in the D4 transition (Fig. 37). Both of these phenomena were observed simultaneously in the R1 transition (Fig. 15). This looks like the splitting of the states in the crystal field that is not possible for the Kramers doublets. Such phenomena can be accounted for by the absorption of the $Nd^{3+}$ ion in two nonequivalent positions. In the former case these can be domains and domain walls and in the latter case: equivalent inverse twins, on the one hand, and the boundaries between the twins, on the other hand. The magnetic field increases and reveals this nonequivalence. The resembling phenomenon was observed in $ErAl_3(BO_3)_4$ and $ErFe_3(BO_3)_4$ crystals [45]. In the $ErAl_3(BO_3)_4$ the local symmetry of $Er^{3+}$ ion ($^4I_{15/2}$) is $D_3$, the same as in the studied crystal. A small splitting (0.11 cm$^{-1}$) of one of the absorption lines was observed without the external magnetic field. In the $ErFe_3(BO_3)_4$ (with the lower symmetry) the same splitting was much larger (11.3 cm$^{-1}$).

### 3.2.6. $^4I_{9/2}\rightarrow{}^2K_{13/2}+{}^4G_{7/2}+{}^4G_{9/2}$ transitions (E+F band)

The total spectrum of this transition was shown in Fig. 10. Majority of lines of this absorption band do not reveal a dependence on temperature and magnetic field. Only lines F1, F2 and F6 demonstrate a substantial dependence on the magnetic field $H||C_3$. The F1 and F2 states have $E_{1/2}$ symmetry (Table 3) and, consequently, they can have $g_\perp\neq 0$ and the exchange splitting. In the F1 state it is ~3.7 cm$^{-1}$ (Fig. 38, Table 3). Notations of the transitions are similar those used above. The F1 and F2 lines reveal no additional splitting in the magnetic field $H\perp C_3$, i. e., this field does not change antiferromagnetic orientation of the $Nd^{3+}$ ion magnetic moments in the F1 and F2 states. The behavior of the F1a$\sigma$ and F1c$\sigma$ line energies versus the magnetic field $H||C_3$ (Fig. 38) was obtained by the decomposition of the spectra into the Gauss components. The opposite and linear dependences of the doublet energies at $H_{||}$>20 kOe testify that in this region magnetic moments of the states change their orientation to the $H||C_3$ direction. The line intensities behavior (Fig. 39, inset) confirms the magnetic moments reorientation. The splitting 26.1 cm$^{-1}$ at $H$=65 kOe corresponds to $g$=8.62 that is close to the theoretical values of $g_C$ and $g_{CM}$ (Table 3).

The exchange splitting of the F2$\pi$ line is not observed. May be, the small difference of the F2$\sigma$ and F2$\pi$ lines energies at $H$=0 (Fig. 40) gives the exchange splitting of the F2 state. The field dependences of the energies and intensities of the F2aπ and F2cπ lines (Fig. 40) show that in the region of 30 – 50 kOe a reorientation of the magnetic moments, to some extent similar to that in the F1 state, takes place.

The F6 state has $E_{3/2}$ symmetry and $g_\perp=0$ (Table 3). Consequently, the exchange splitting of the state should be zero, and the magnetic field $H\perp C_3$ also does not split the corresponding transition. Since there is no the exchange splitting of the F6 state, the ion is actually in the paramagnetic state. Therefore it is rather strange that the splitting in the $H\|C_3$ is not proportional to the magnetic field (Fig. 41), in contrast to that observed in the S8 state (Fig. 27), which also has $g_\perp=0$. The F6cσ line appears only at $H_\|=40$ kOe. Therefore, it is impossible to retrace the F6cσ transition energy behavior in the lower fields. The kink at 30 kOe in the F6aσ line energy behavior and the linear dependence on the magnetic field apparently testify to reorientation of magnetic moments to the $H\|C_3$ direction.

## 4. Summary

Polarized absorption spectra and magnetic circular dichroism MCD spectra of $Nd_{0.5}Gd_{0.5}Fe_3(BO_3)_4$ single crystal were measured in the range of 10000 – 21000 cm$^{-1}$. The absorption spectra were studied as a function of temperature in the range of 2 – 293 K and as a function of magnetic field 0 – 65 kOe at 2 K. In the paramagnetic state of the crystal, the excited $4f$ states of the $Nd^{3+}$ ion were identified in terms of the irreducible representations and in terms of the $|J,\pm M_J\rangle$ wave functions of the free atom. Basing on this wave function approximations, the Landé factors of the $4f$ states and the changes of the Landé factor during the $f$-$f$ transitions were calculated. With the help of the absorption and MCD spectra the changes of the Landé factor during a number of the $f$-$f$ transitions were found experimentally. In the majority of cases the experimentally found values were close to the theoretically predicted ones. The discrepancies, observed in some cases, can be accounted for by the mixing of the $|J,\pm M_J\rangle$ states with the different $M_J$ but with the same crystal quantum number $\mu$.

In the magnetically ordered state of the crystal, the exchange splitting of the ground and excited states were determined. The exchange field of the Fe sublattice is directed perpendicular to the $C_3$ axis. Therefore, the exchange splitting of the excited $Nd^{3+}$ states with $g_\perp=0$ should be zero. It is really so (see Table 3). However, in some excited states the $Nd^{3+}$ ions revealed the zero exchange splitting, i. e., the zero exchange interaction with the Fe sublattice, in spite of the not zero Landé factor g$_\perp$. Thus, the value of the exchange splitting (the exchange interaction) does not correlate with the theoretical Landé factor g$_\perp$ (Table 3), and in general, the local magnetic properties in the vicinity of the excited ion substantially depend on its electron state. In particular: 1) in some excited states a weak ferromagnetic moment appears in the easy plane or in the $C_3$ direction, 2) the reorientation field induced magnetic transitions occur, 3) the changes of type of the local magnetic anisotropy take place.

It was found out that the selection rules for the electron transitions substantially changed in the magnetically ordered state of the crystal. They are different for the different transitions and they strongly depend on the orientation of the sublattice magnetic moment relative to the light polarization.

In some cases, the energies of the transitions depend on polarization. Such phenomenon can be accounted for by the absorption of the $Nd^{3+}$ ions in two nonequivalent positions. In the many domain state ($H_\perp$<4 kOe) these can be domains and domain walls. In the one domain state ($H_\perp$>4 kOe) these can be equivalent inverse twins, on the one hand, and the boundaries between the twins, on the other hand. The magnetic field increases and reveals this nonequivalence. This phenomenon is observed only in some excited states. Consequently, it is connected with the electron state dependent local distortions.

It is worth noting that majority of the studied excited states has identical symmetry $E_{1/2}$ but a great variety of the local magnetic properties and of the polarization properties of the corresponding transitions. There can be two sources of these features. First, these states refer to different multiplets and (or) have different $M_J$ in the $|J,\pm M_J\rangle$ representation. Second, in spite of the axial crystal symmetry, the magnetic moments in the ground state are in the plane perpendicular to the $C_3$ axis. Therefore, they create one more quantization axis besides the $C_3$ one, and, consequently, they create the peculiar selection rules for the electron transitions, which depend also on the excited states because the electronically excited atom modifies the local symmetry and magnetic properties of the crystal.

**Acknowledgements**

The work was supported by the President of Russia grant No Nsh-2886.2014.2.

**Figure captions**

Fig. 1. Polarized absorption spectra of the *f-f* transitions in $Nd_{0.5}Gd_{0.5}Fe_3(BO_3)_4$ crystal at room temperature. The final states of the $Nd^{3+}$ ion are indicated.

Fig. 2. General scheme of *f-f* transitions in $Nd^{3+}$ ion.

Fig. 3. Polarized absorption spectra (a) and MCD spectra (b) of the $^4I_{9/2}\rightarrow{}^4F_{3/2}$ transition (R-band) at $T$=90 K.

Fig. 4. Polarized absorption spectra of the $^4I_{9/2}\rightarrow{}^2H_{9/2}+{}^4F_{5/2}$ transitions (S-band) at $T$=6 K.

Fig. 5. Polarized absorption spectra (a), the first derivative of MCD (b) and MCD spectra (c) of the $^4I_{9/2}\rightarrow{}^2H_{9/2}+{}^4F_{5/2}$ transitions (S-band) at $T$=90 K.

Fig. 6. Polarized absorption spectra of the $^4I_{9/2}\rightarrow{}^4S_{3/2}+{}^4F_{7/2}$ transitions (A-band) at $T$=2 K.

Fig. 7. Polarized absorption spectra (a) and MCD spectra (b) of the $^4I_{9/2}\rightarrow{}^4S_{3/2}+{}^4F_{7/2}$ transitions (A-band) at $T$=90 K.

Fig. 8. Polarized absorption spectra of the $^4I_{9/2}\rightarrow{}^4G_{5/2}+{}^2G_{7/2}$ transitions (D-band) at $T$=6 K.

Fig. 9. Polarized absorption spectra (a), the first derivative of MCD (b) and MCD spectra (c) of the $^4I_{9/2}\rightarrow{}^4G_{5/2}+{}^2G_{7/2}$ transitions (D-band) at $T$=90 K.

Fig. 10. Polarized absorption spectra of the $^4I_{9/2}\rightarrow{}^4G_{9/2}+{}^4G_{7/2}+{}^2K_{13/2}$ transitions (E+F bands) at $T$=2 K.

Fig. 11. Polarized absorption spectra (a), the first derivative of MCD (b) and MCD spectra (c) of the $^4I_{9/2}\rightarrow{}^4G_{9/2}+{}^4G_{7/2}+{}^2K_{13/2}$ transitions (E+F bands) at $T$=90 K.

Fig. 12. Temperature dependences of the exchange splitting of transitions into the states of the $E_{3/2}$ symmetry.

Fig. 13. Geometry of experiments.

Fig. 14. Diagram of the R1 transitions.

Fig. 15. Energies of the R1a line in two polarizations as a function of the magnetic field $H\perp C_3$. Inset: intensities of the same lines.

Fig. 16. Diagram of the S2 transitions.

Fig. 17. Absorption spectra of the S2π transition in the magnetic field $H||C_3$.

Fig. 18. Energies of the S2aπ and S2cπ transitions as a function of the magnetic field $H\perp C_3$ and $H||C_3$. Inset: Splitting of the S2 line in the magnetic field $H||C_3$.

Fig. 19. Absorption spectra of the S4σ transition in the magnetic fields $H\perp C_3$ and $H||C_3$.

Fig. 20. Energies of the S4aπ, S4aσ and S4cσ transitions as a function of the magnetic field $H\perp C_3$. Inset: field dependences of intensities of the S4σ transitions.

Fig. 21. Diagram of the S4 transitions.

Fig. 22. Energies of the S4 transitions as a function of magnetic field $H||C_3$.

Fig. 23. Temperature dependence of the exchange splitting of the S4σ transition. Inset: absorption spectrum of the S4σ transition at several temperatures.

Fig. 24. Absorption spectrum of the S5σ and S6σ transitions in the magnetic field $H_\perp$=65 kOe. Inset: absorption spectra of the S5π transition in two magnetic fields $H\perp C_3$.

Fig. 25. Energies of the S5 transitions as a function of the magnetic field $H\perp C_3$ and $H\|C_3$.

Fig. 26. Energies of the S7 transitions as a function of the magnetic field $H\perp C_3$. Inset: absorption spectra of the S7π transition in two magnetic fields $H\perp C_3$.

Fig. 27. Energies of the S8 transitions as a function of the magnetic field $H\perp C_3$ and $H\|C_3$.

Fig. 28. Energies of the A1 transitions as a function of the magnetic field $H\perp C_3$.

Fig. 29. Diagram of the A1 transitions.

Fig. 30. Temperature dependence of the exchange splitting of the A5π transition.

Fig. 31. Diagram of the A5 transitions.

Fig. 32. Energies of the A5 and A6 transitions as a function of the magnetic field $H\perp C_3$. Inset: fragment of the A band spectrum at several magnetic fields $H\perp C_3$.

Fig. 33. Diagram of the D1 transitions.

Fig. 34. Intensities of the D1aσ and D1cσ lines as a function of the magnetic field $H\perp C_3$.

Fig. 35. Energies (*E*) and intensities (*I*) of the D1 lines as a function of temperature.

Fig. 36. Intensities of the D5aπ and D5cπ lines as a function of the magnetic field $H\|C_3$. Inset: transformation of the D5π transition spectrum in the magnetic field $H\|C_3$.

Fig. 37. Energies of the D1a and D4a transitions in two polarizations as a function of the magnetic field $H\perp C_3$. Inset: Splitting of the D2 line in the magnetic field $H\perp C_3$.

Fig. 38. Energies of the F1aσ and F1cσ transitions as a function of the magnetic field $H\|C_3$. Inset: transformation of the F1σ spectrum in the magnetic field $H\|C_3$.

Fig. 39. Line widths and intensities (inset) of F1aσ and F1cσ transitions as a function of the magnetic field $H\|C_3$.

Fig. 40. Energies and intensities (inset) of the F2aπ and F2cπ transitions as a function of the magnetic field $H\|C_3$.

Fig. 41. Energies and intensities (inset) of the F6aσ and F6cσ transitions as a function of the magnetic field $H\|C_3$.

Table 1. Selection rules for electric dipole transitions in $D_3$ symmetry

|  | $E_{1/2}$ | $E_{3/2}$ |
|---|---|---|
| $E_{1/2}$ | $\pi, \sigma(\alpha)$ | $\sigma(\alpha)$ |
| $E_{3/2}$ | $\sigma(\alpha)$ | $\pi$ |

Table 2. Landé factors of the Kramers doublets along the $C_3$ axis of a crystal in approximation of the $|J,\pm M_J\rangle$ functions.

|  | M | 13/2 | 11/2 | 9/2 | 7/2 | 5/2 | 3/2 | 1/2 |
|---|---|---|---|---|---|---|---|---|
| $^4I_{9/2}$, g=0.727 | $g_{CM}$ |  |  | 6.54 | 5.09 | 3.64 | 2.18 | 0.727 |
| $^4F_{3/2}$, g=0.4 | $g_{CM}$ |  |  |  |  |  | 1.2 | 0.4 |
| $^4F_{5/2}$, g=1.029 | $g_{CM}$ |  |  |  |  | 5.14 | 3.09 | 1.03 |
| $^2H_{9/2}$, g=0.909 | $g_{CM}$ |  |  | 8.18 | 6.36 | 4.55 | 2.73 | 0.909 |
| $^4S_{3/2}$, g=2 | $g_{CM}$ |  |  |  |  |  | 6 | 2 |
| $^4F_{7/2}$, g=1.238 | $g_{CM}$ |  |  |  | 8.67 | 6,19 | 3.71 | 1.238 |
| $^4G_{5/2}$, g=0.571 | $g_{CM}$ |  |  |  |  | 2.855 | 1.713 | 0.571 |
| $^2G_{7/2}$, g=0.889 | $g_{CM}$ |  |  |  | 6.223 | 4.445 | 2.667 | 0.889 |
| $^4G_{7/2}$, g=0.984 | $g_{CM}$ |  |  |  | 6.888 | 4.92 | 2.952 | 0.984 |
| $^4G_{9/2}$, g=1.1717 | $g_{CM}$ |  |  | 10.54 | 8.202 | 5.86 | 3.515 | 1.1717 |
| $^2K_{13/2}$, g=0.933 | $g_{CM}$ | 12.1 | 10.27 | 8,4 | 6.53 | 4.66 | 2.8 | 0.933 |

Table 3. Parameters of transitions and states. Δ$E1$ – exchange splitting of absorption lines connected with the exchange splitting of the ground state. Δ$E2$ – experimentally observed exchange splitting of excited states. Energies of transitions ($E$) are given at 40 K. (Details are in the text).

| State | Level | $E$ cm$^{-1}$ | Polar. | Sym. | $\mu$ | $M_J$ | Δ$g_C$ (exp) | Δ$g_{CM}$ | Δ$E1$ cm$^{-1}$ ($\pi$) | Δ$E1$ cm$^{-1}$ ($\sigma$) | Δ$E2$ cm$^{-1}$ | $g_\perp$ [22] | $g_C$ [22] | $g_{CM}$ |
|---|---|---|---|---|---|---|---|---|---|---|---|---|---|---|
| $^4I_{9/2}$ | Gr1 | 0 | - | $E_{1/2}$ | ∓1/2 | ±5/2 | | | | | | 2.385 | 1.376 | 3.64 |
| | Gr2 | ~78 | - | $E_{3/2}$ | 3/2 | ±9/2 | | | | | | 0 | 3.947 | 6.54 |
| | Gr3 | 150 | - | $E_{1/2}$ | ±1/2 | ±7/2 | | | | | | 2.283 | 2.786 | 5.09 |
| | Gr4 | 220 | - | $E_{3/2}$ | 3/2 | ±3/2 | | | | | | 0 | 3.879 | 2.18 |
| | Gr5 | | - | $E_{1/2}$ | ±1/2 | ±1/2 | | | | | | 3.527 | 0.843 | 0.727 |
| $^4F_{3/2}$ | R1 | 11369 | $\pi, \sigma$ | $E_{1/2}$ | ±1/2 | ±1/2 | (+) | +3.24 | | | | 0.926 | 0.251 | 0.4 |
| | R2 | 11436 | $\sigma$ | $E_{3/2}$ | 3/2 | ±3/2 | -3.4 | -4.84 | | 10 | 0 | 0 | 1.562 | 1.2 |
| $^4F_{5/2}$ | S1 | 12382 | $\pi, \sigma$ | $E_{1/2}$ | ±1/2 | ±1/2 | (-) | +2.61 | | | | 3.158 | 0.598 | 1.03 |
| | S2 | 12450 | $\pi, \sigma$ | $E_{1/2}$ | ∓1/2 | ±5/2 | +8.4 | +8.78 | 9.6 | | 2.7 | 0.096 | 4.713 | 5.14 |
| | S3 | 12467 | $\sigma$ | $E_{3/2}$ | 3/2 | ±3/2 | (-) | -6.73 | | | | 0 | 2.576 | 3.09 |
| $^2H_{9/2}$ | S4 | 12495 | $\pi, \sigma$ | $E_{1/2}$ | ±1/2 | ±7/2 | -3.3 | -2.72 | 8 | 9 | 0 | 1.989 | 4.633 | 6.36 |
| | S5 | 12553 | $\pi, \sigma$ | $E_{1/2}$ | ±1/2 | ±1/2 | | +2.73 | | | 0 | 3.995 | 0.982 | 0.909 |
| | S6 | 12578 | $\sigma$ | $E_{3/2}$ | 3/2 | ±3/2 | | -6.37 | | | 0 | 0 | 2.169 | 2.73 |
| | S7 | 12620 | $\pi, \sigma$ | $E_{1/2}$ | ∓1/2 | ±5/2 | +4.9 | +8.19 | | | 0 | 2.877 | 2.765 | 4.55 |
| | S8 | 12702 | $\sigma$ | $E_{3/2}$ | 3/2 | ±9/2 | +13.9 | -11.8 | | 8.8 | 0 | 0 | 7.788 | 8.18 |
| $^4F_{7/2}$ | A1 | 13353 | $\pi, \sigma$ | $E_{1/2}$ | ±1/2 | ±1/2 | | +2.4 | | | ~0.5 | 3.216 | 0.673 | 1.238 |
| | A2 | 13370 | $\sigma$ | $E_{3/2}$ | 3/2 | ±3/2 | -8.6 | -7.35 | | | | 0 | 3.524 | 3.71 |
| | A3 | | $\pi,(\sigma?)$ | $E_{1/2}$ | ∓1/2 | ±5/2 | | +9.83 | | | | 4.895 | 1.296 | 6.19 |
| | A4 | | $(\pi?)\sigma$ | $E_{1/2}$ | ±1/2 | ±7/2 | | -5.03 | | | | 3.858 | 2.141 | 8.67 |
| $^4S_{3/2}$ | A5 | 13488 | $\pi, \sigma$ | $E_{1/2}$ | ±1/2 | ±1/2 | | +1.64 | 16.4 | | 7.0 | 3.239 | 1.884 | 2 |
| | A6 | 13499 | $\sigma$ | $E_{3/2}$ | 3/2 | ±3/2 | -7.55 | -9.64 | | | | 0 | 5.848 | 6 |

| | | | | | | | | | | | | | |
|---|---|---|---|---|---|---|---|---|---|---|---|---|---|
| $^4G_{5/2}$ | D1 | 16921 | $\pi, \sigma$ | $E_{1/2}$ | ±1/2 | ±1/2 | (+) | +3.07 | | | 7 | 0.043 | 0.065 | 0.571 |
| | D2 | 17062 | $\pi, \sigma$ | $E_{1/2}$ | ∓1/2 | ±5/2 | (+) | +6.5 | 14.2 | | 1 | 1.385 | 1.310 | 2.885 |
| | D3 | 17100 | $\approx \sigma$ | $E_{3/2}$ | 3/2 | ±3/2 | (-) | -5.35 | | | 0 | 0 | 3.044 | 1.713 |
| $^2G_{7/2}$ | D4 | 17199 | $\pi, \sigma$ | $E_{1/2}$ | ±1/2 | ±1/2 | (-) | +2.75 | | | 12 | 2.617 | 0.266 | 0.889 |
| | D5 | 17240 | $\pi, \sigma$ | $E_{1/2}$ | ±1/2 | ±7/2 | (+) | -2.58 | ~17 | | 7.5 | 0.755 | 3.308 | 6.223 |
| | D6 | 17289 | $\pi, \sigma$ | $E_{1/2}$ | ∓1/2 | ±5/2 | (-) | +8.08 | | | 0.4 | 1.538 | 0.954 | 4.445 |
| | D7 | 17325 | $\approx \sigma$ | $E_{3/2}$ | 3/2 | ±3/2 | (-) | -6.31 | | | 0 | 0 | 1.016 | 2.667 |
| | | | | | | | | | | | | | | |
| $^4G_{9/2}$ | E1 | 18872 | $\pi, \sigma$ | $E_{1/2}$ | ±1/2 | ±7/2 | (-) | +2.47 | | | | 3.369 | 1.784 | 1.17 |
| | E2 | 18910 | $\pi, \sigma$ | $E_{1/2}$ | ∓1/2 | ±5/2 | (-) | +9.5 | | | | 2.098 | 3.026 | 5.86 |
| | E3 | 18959 | $\approx \sigma$ | $E_{3/2}$ | 3/2 | ±3/2 | (-) | -7.17 | | | | 0 | 2.924 | 3.515 |
| | E4 | 19010 | $\pi, \sigma$ | $E_{1/2}$ | ±1/2 | ±1/2 | | -4.56 | | | | 2.292 | 2.081 | 8.202 |
| | E5 | 19064 | $\sigma$ | $E_{3/2}$ | 3/2 | ±9/2 | | -14.2 | | | | 0 | 8.530 | 10.54 |
| | | | | | | | | | | | | | | |
| $^2K_{13/2}$ | F1 | 19295 | $\pi, \sigma$ | $E_{1/2}$ | ∓1/2 | ±11/2 | +10.7 | +13.9 | | | 3.7 | 0.283 | 9.760 | 10.27 |
| $+^4G_{7/2}$ | F2 | 19323 | $\pi, \sigma$ | $E_{1/2}$ | | | | | | | ~1 | | | |
| | F3 | 19389 | $\pi, \sigma$ | $E_{1/2}$ | | | | | | | | | | |
| | F4 | 19408 | $\pi, \sigma$ | $E_{1/2}$ | | | | | | | | | | |
| | F5 | 19438 | $\pi, \sigma$ | $E_{1/2}$ | | | | | | | | | | |
| | F6 | 19461 | $\sigma$ | $E_{3/2}$ | 3/2 | ±9/2 | +8.8 | -12.0 | | | 0 | 0 | 7.030 | 8.4 |
| | F7 | 19525 | $\pi, \sigma$ | $E_{1/2}$ | | | (-) | | | | | | | |
| | F8 | 19687 | $\pi, \sigma$ | $E_{1/2}$ | | | | | | | | | | |
| | F9 | 19769 | $\sigma$ | $E_{3/2}$ | 3/2 | ±3/2(G) | (-) | -6.6 | | | | 0 | 3.025 | 2.952 |
| | F10 | 19793 | $\pi, \sigma$ | $E_{1/2}$ | | | | | | | | | | |

Fig. 1. Polarized absorption spectra of the f-f transitions in $Nd_{0.5}Gd_{0.5}Fe_3(BO_3)_4$ crystal at room temperature. The final states of the $Nd^{3+}$ ion are indicated.

Fig. 2. General scheme of f-f transitions in $Nd^{3+}$ ion.

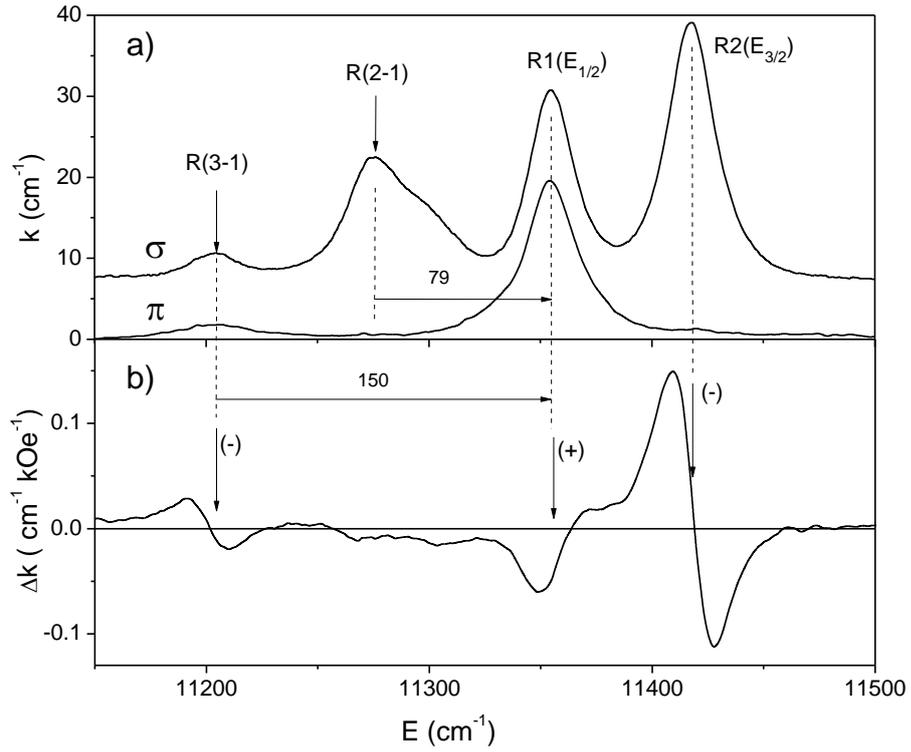

Fig. 3. Polarized absorption spectra (a) and MCD spectra (b) of the $^4I_{9/2} \rightarrow {}^4F_{3/2}$ transition (R-band) at $T$=90 K.

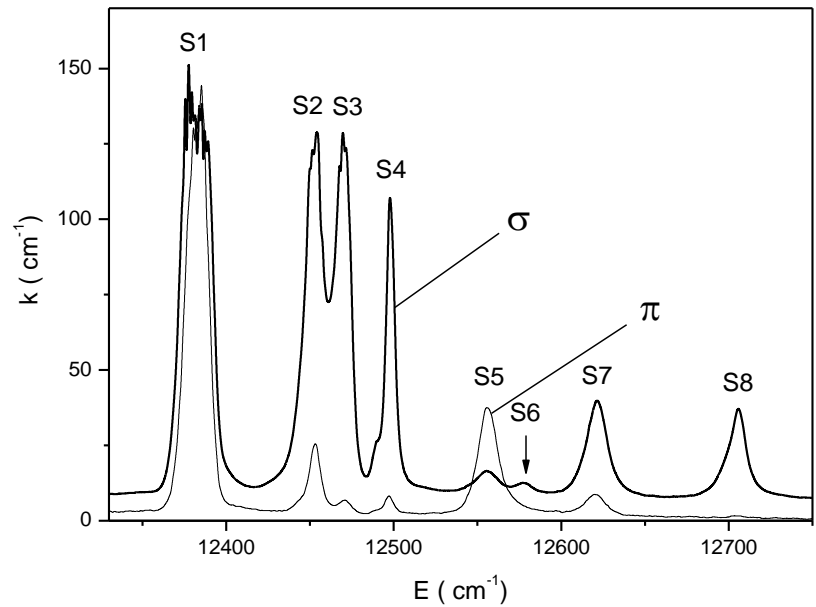

Fig. 4. Polarized absorption spectra of the $^4I_{9/2} \rightarrow {}^2H_{9/2} + {}^4F_{5/2}$ transitions (S-band) at $T$=6 K.

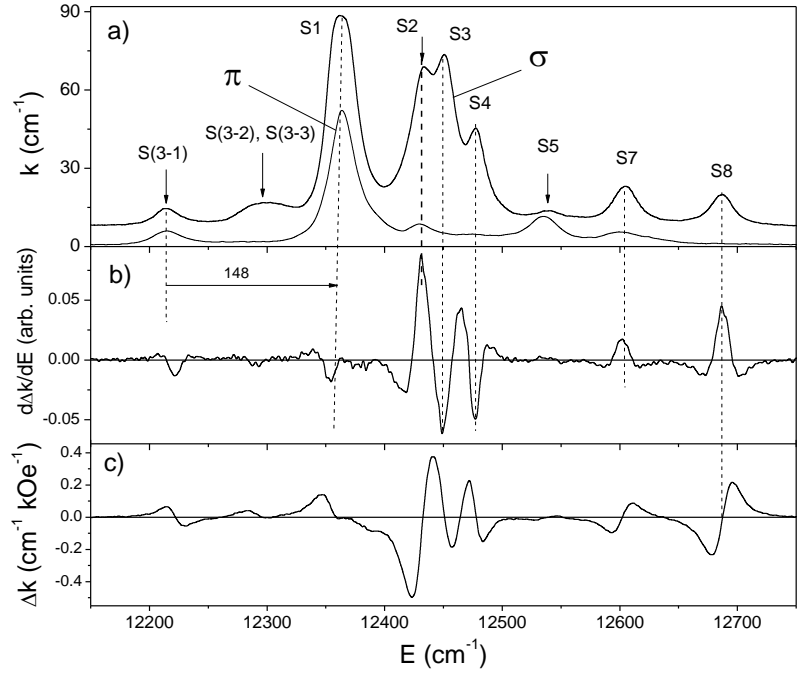

Fig. 5. Polarized absorption spectra (a), the first derivative of MCD (b) and MCD spectra (c) of the $^4I_{9/2} \to {}^2H_{9/2} + {}^4F_{5/2}$ transitions (S-band) at $T$=90 K.

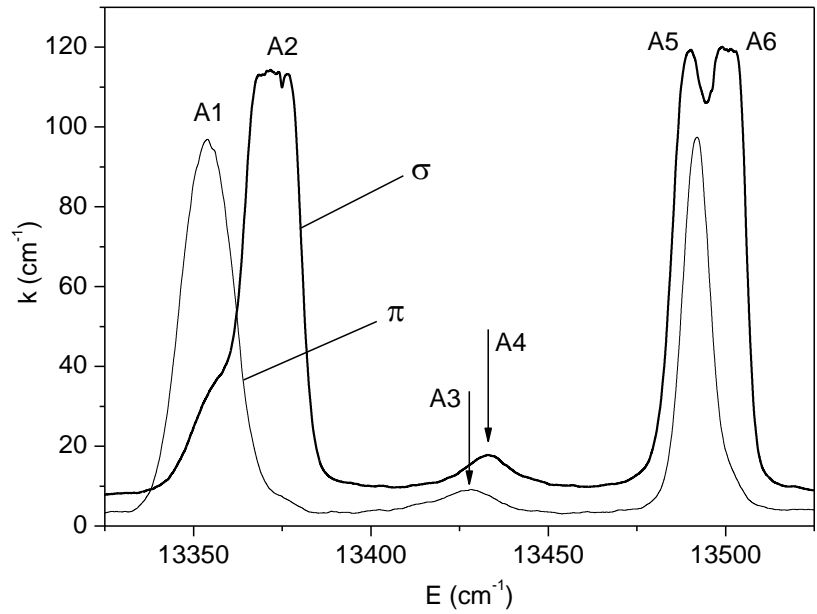

Fig. 6. Polarized absorption spectra of the $^4I_{9/2} \to {}^4S_{3/2} + {}^4F_{7/2}$ transitions (A-band) at $T$=2 K.

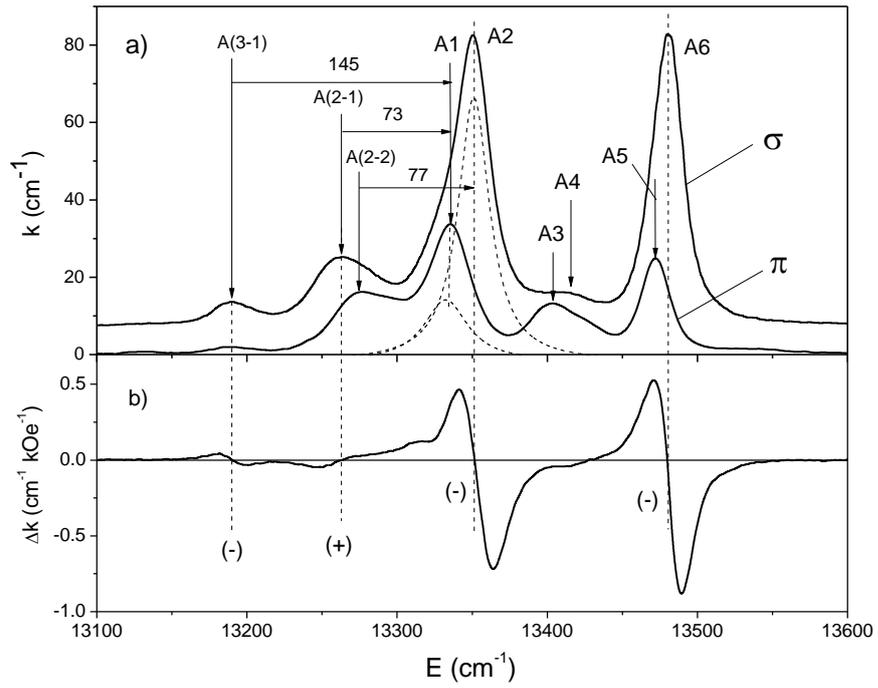

Fig. 7. Polarized absorption spectra (a) and MCD spectra (b) of the $^4I_{9/2} \rightarrow {}^4S_{3/2} + {}^4F_{7/2}$ transitions (A-band) at $T$=90 K.

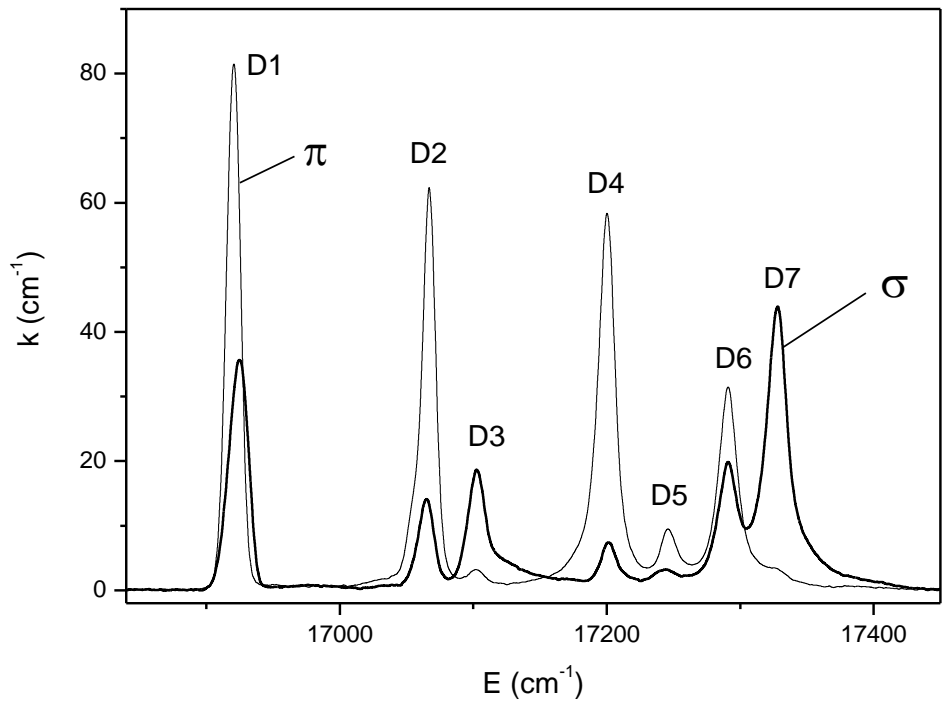

Fig. 8. Polarized absorption spectra of the $^4I_{9/2} \rightarrow {}^4G_{5/2} + {}^2G_{7/2}$ transitions (D-band) at $T$=6 K.

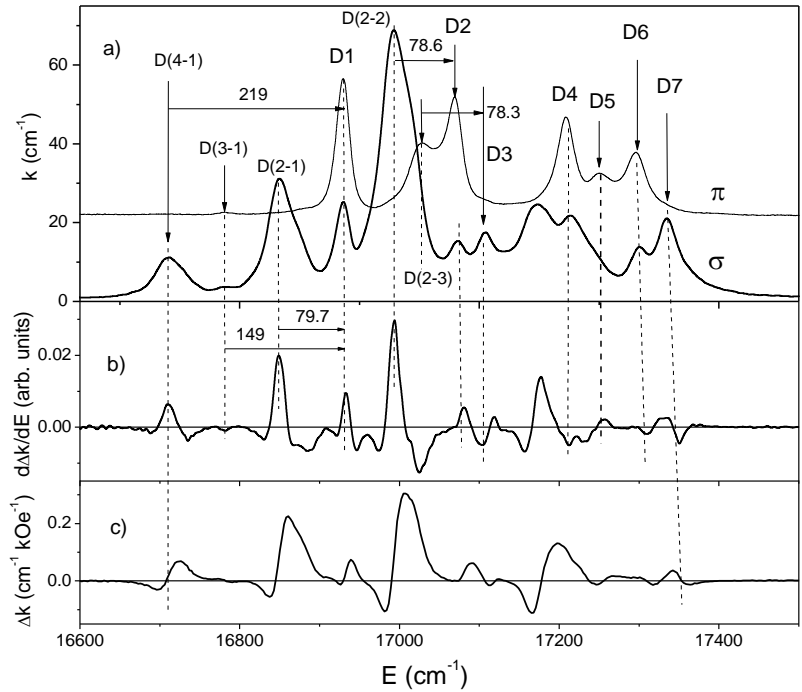

Fig. 9. Polarized absorption spectra (a), the first derivative of MCD (b) and MCD spectra (c) of the $^4I_{9/2} \to {}^4G_{5/2}+{}^2G_{7/2}$ transitions (D-band) at $T$=90 K.

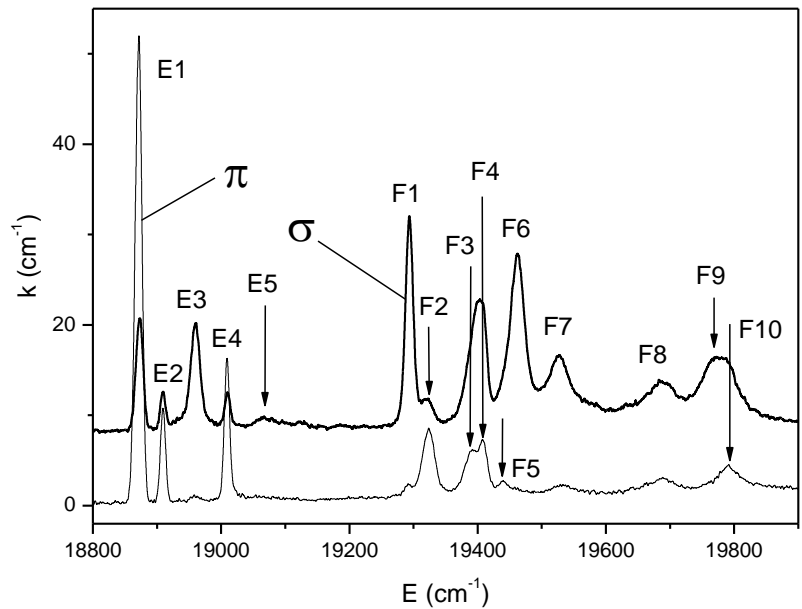

Fig. 10. Polarized absorption spectra of the $^4I_{9/2} \to {}^4G_{9/2}+{}^4G_{7/2}+{}^2K_{13/2}$ transitions (E+F-bands) at $T$=2 K.

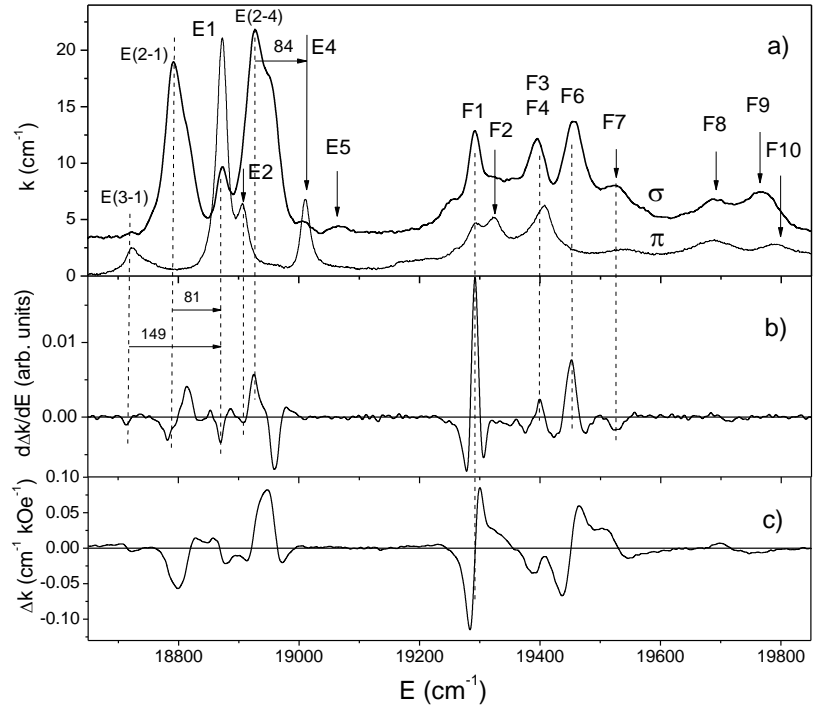

Fig. 11. Polarized absorption spectra (a), the first derivative of MCD (b) and MCD spectra (c) of the $^4I_{9/2} \rightarrow {}^4G_{9/2} + {}^4G_{7/2} + {}^2K_{13/2}$ transitions (E+F bands) at $T$=90 K.

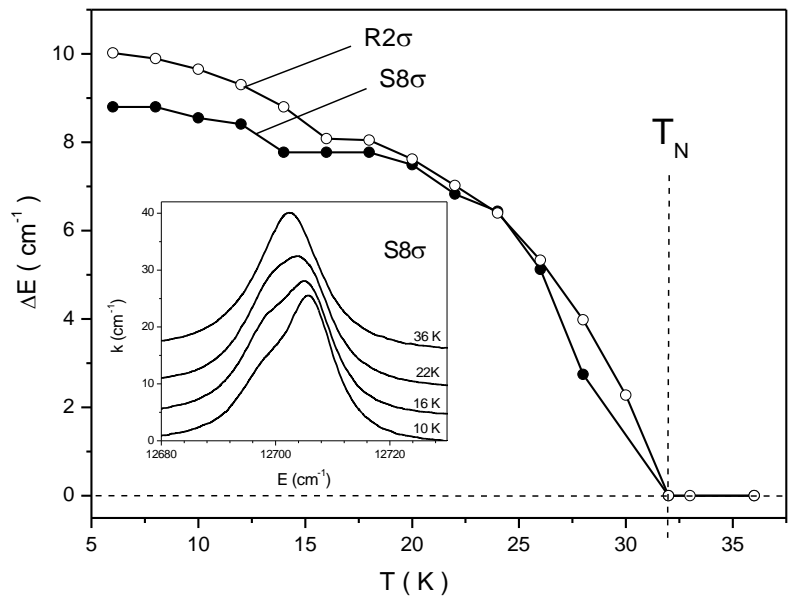

Fig. 12. Temperature dependences of the exchange splitting of transitions into the states of the $E_{3/2}$ symmetry.

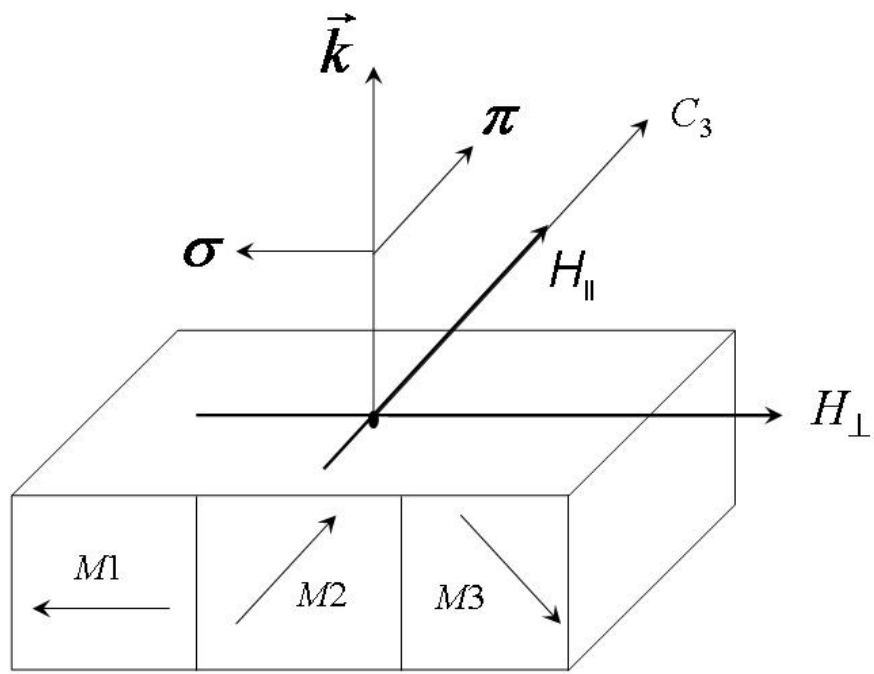

Fig. 13. Geometry of experiments.

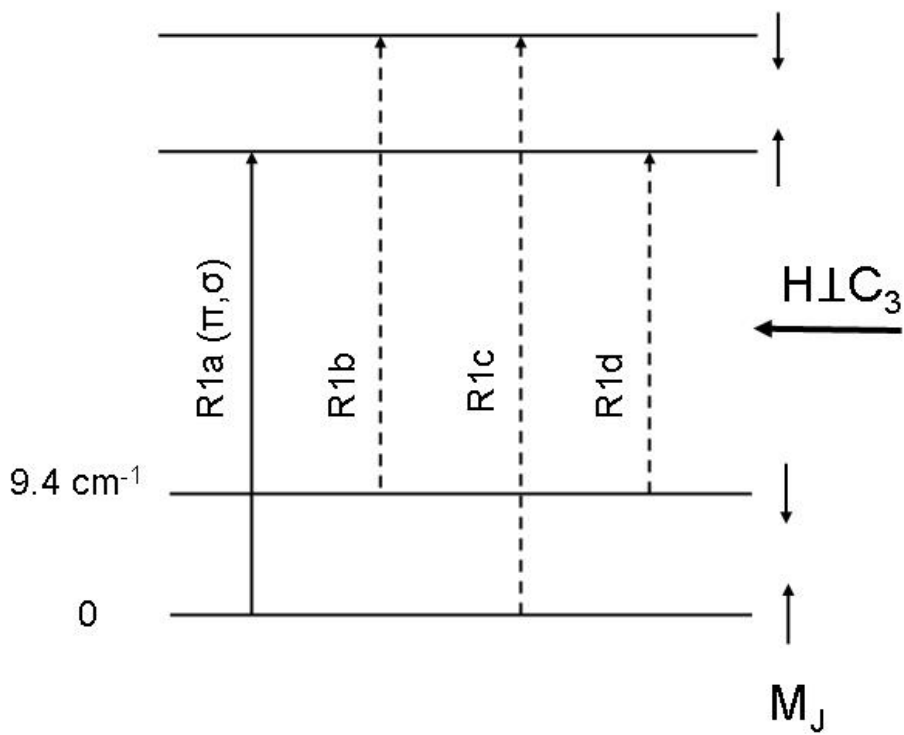

Fig. 14. Diagram of the R1 transitions.

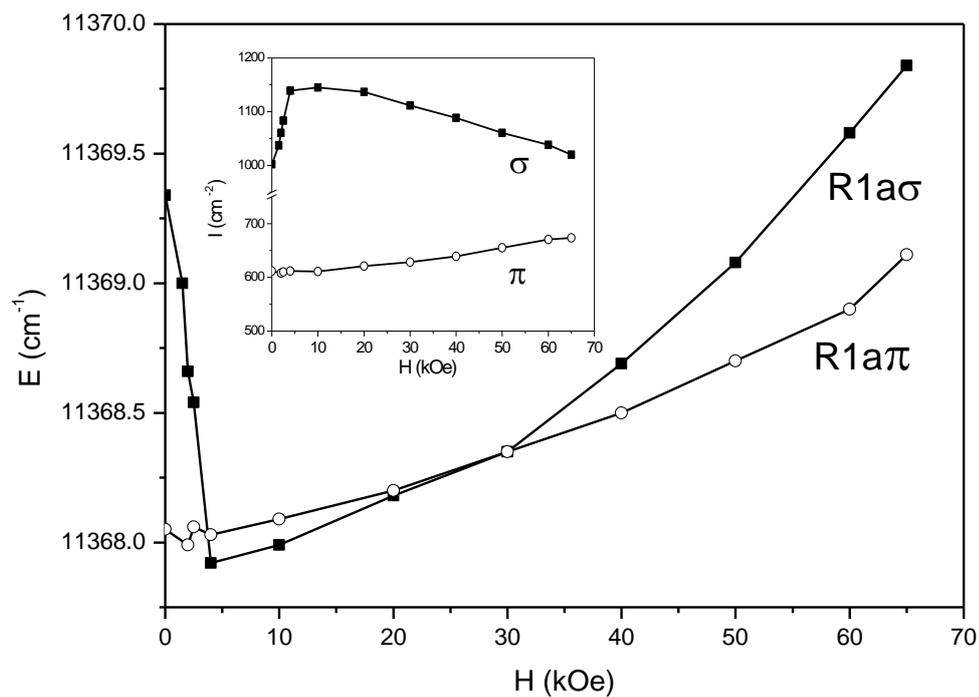

Fig. 15. Energies of the R1a line in two polarizations as a function of the magnetic field $H\perp C_3$. Inset: intensities of the same lines.

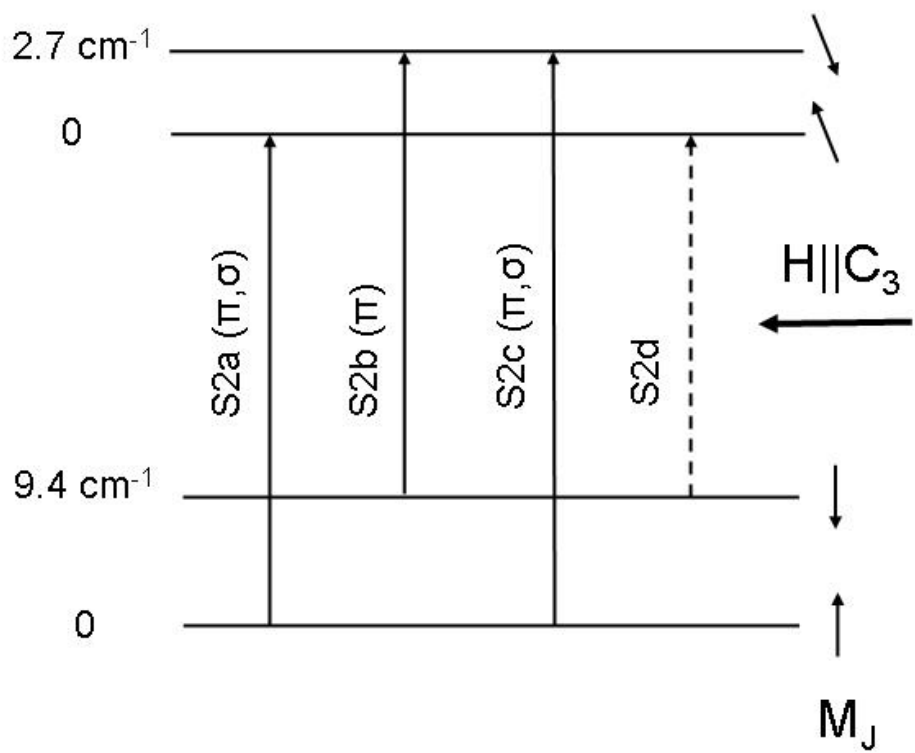

Fig. 16. Diagram of the S2 transitions.

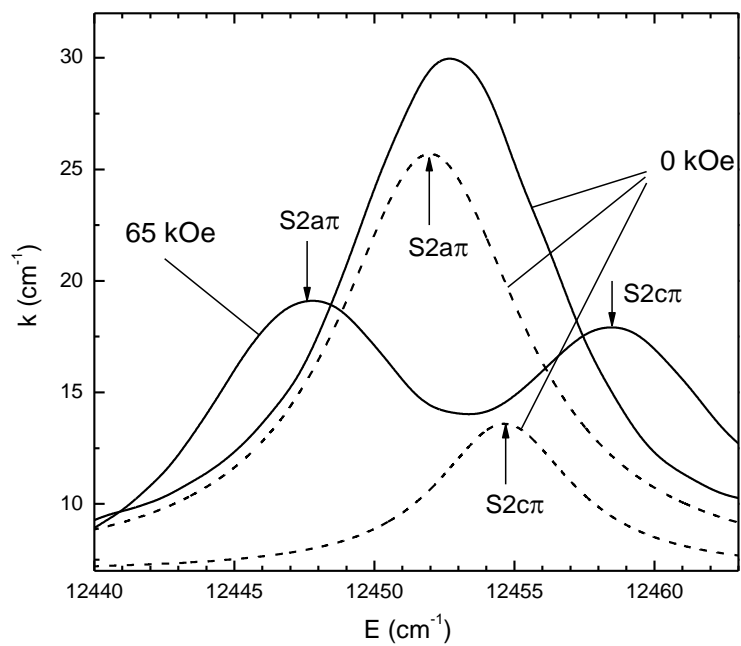

Fig. 17. Absorption spectra of the S2π transition in the magnetic field $H\|C_3$.

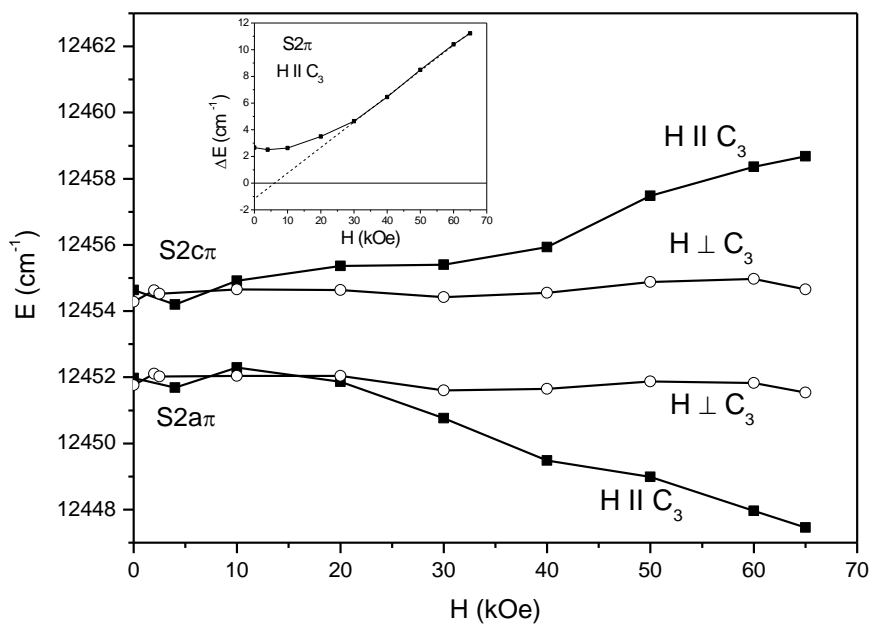

Fig. 18. Energies of the S2aπ and S2cπ transitions as a function of the magnetic field $H\perp C_3$ and $H\|C_3$. Inset: Splitting of the S2 line in the magnetic field $H\|C_3$.

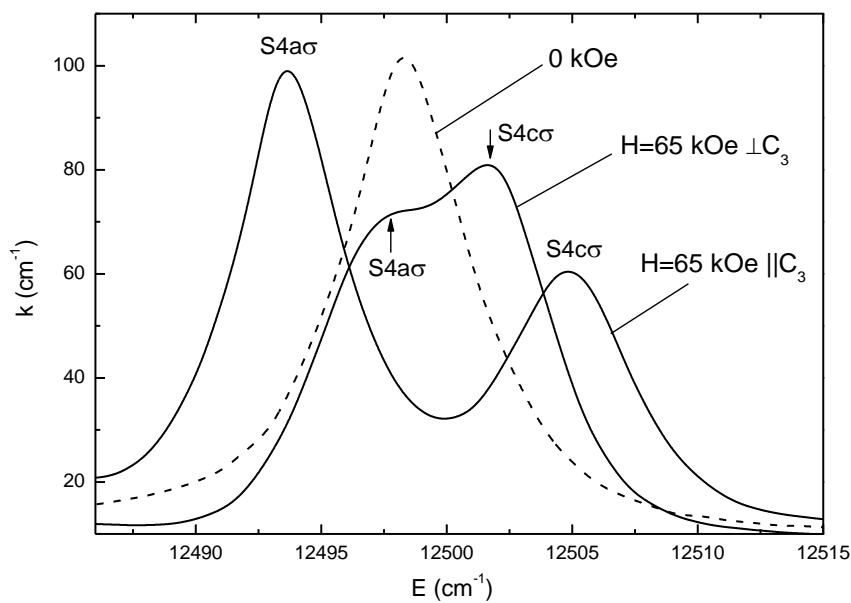

Fig. 19. Absorption spectra of the S4σ transition in the magnetic fields H⊥C$_3$ and H∥C$_3$.

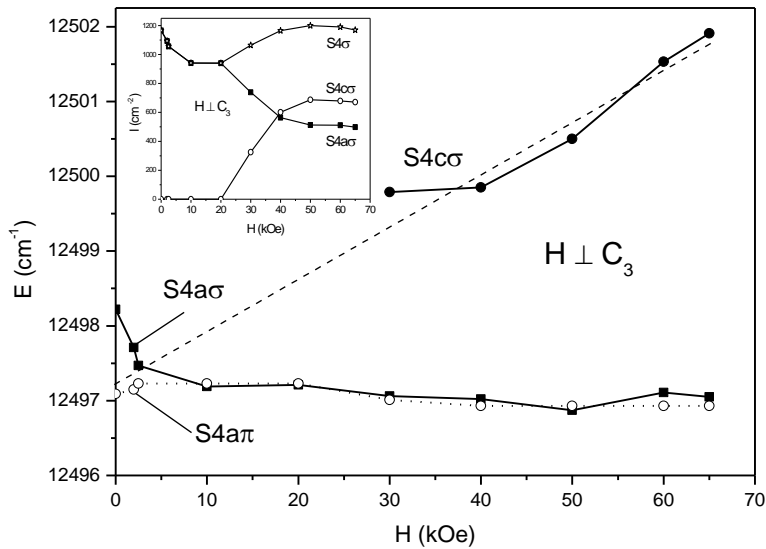

Fig. 20. Energies of the S4aπ, S4aσ and S4cσ transitions as a function of the magnetic field $H \perp C_3$. Inset: field dependences of intensities of the S4σ transitions.

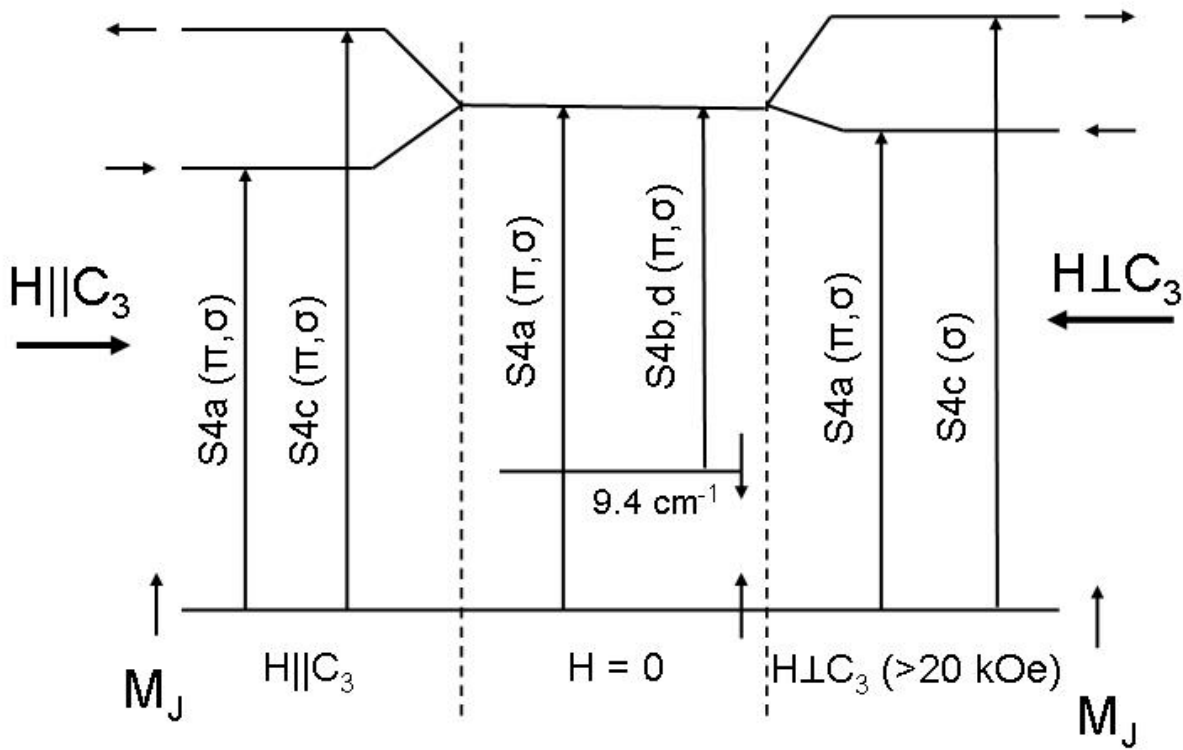

Fig. 21. Diagram of the S4 transitions.

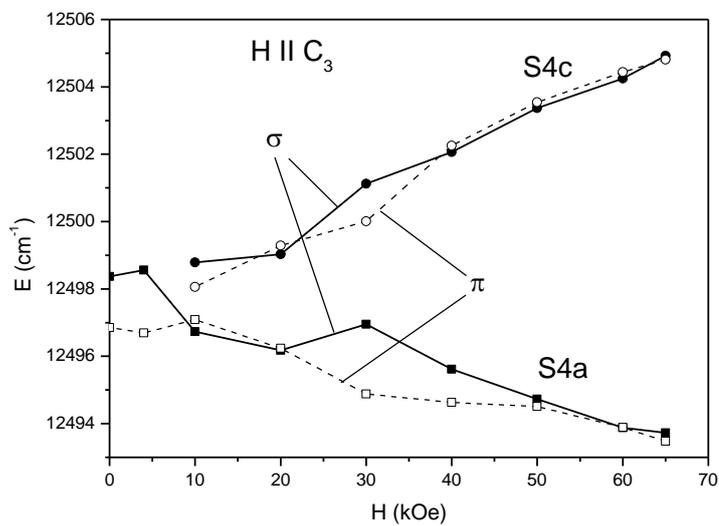

Fig. 22. Energies of the S4 transitions as a function of magnetic field $H \| C_3$.

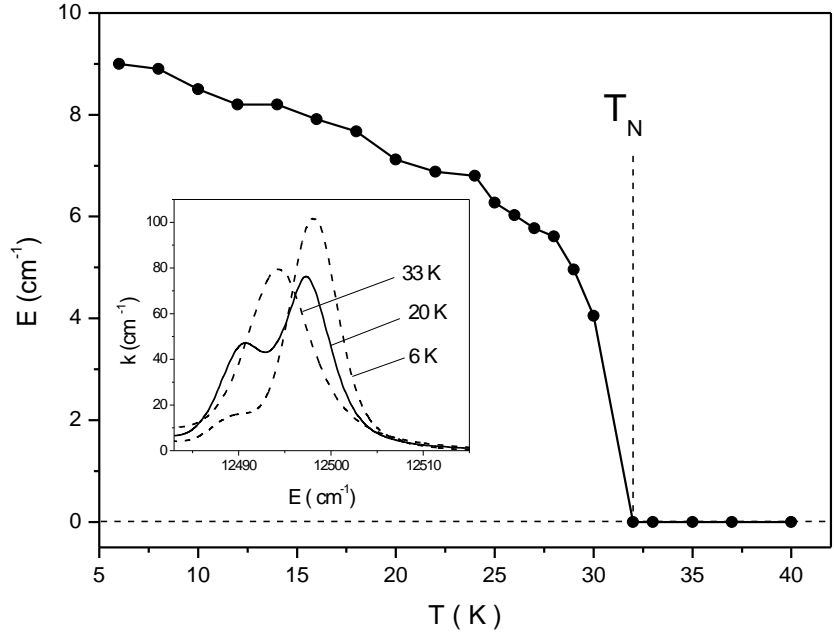

Fig. 23. Temperature dependence of the exchange splitting of the S4σ transition. Inset: absorption spectrum of the S4σ transition at several temperatures.

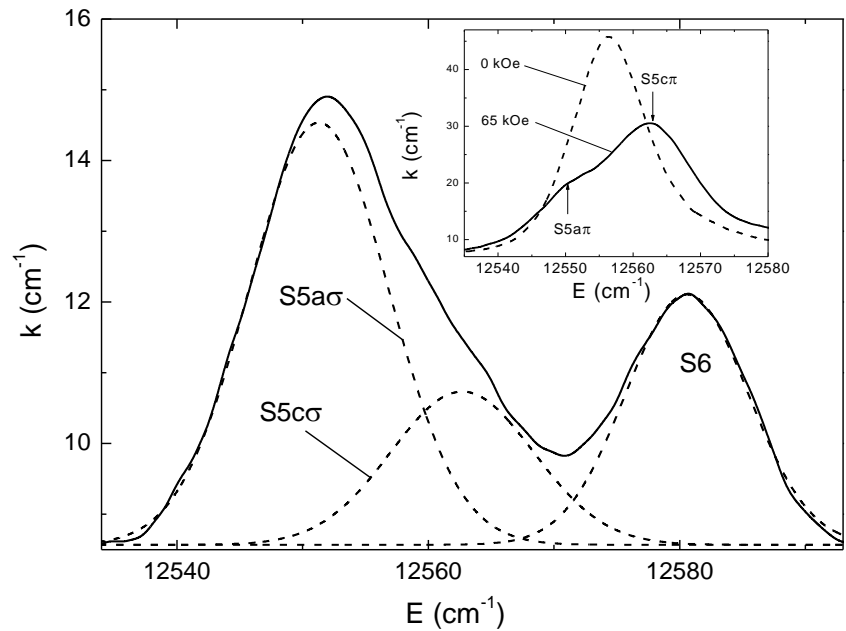

Fig. 24. Absorption spectrum of the S5σ and S6σ transitions in the magnetic field $H_\perp$=65 kOe. Inset: absorption spectra of the S5π transition in two magnetic fields $H \perp C_3$.

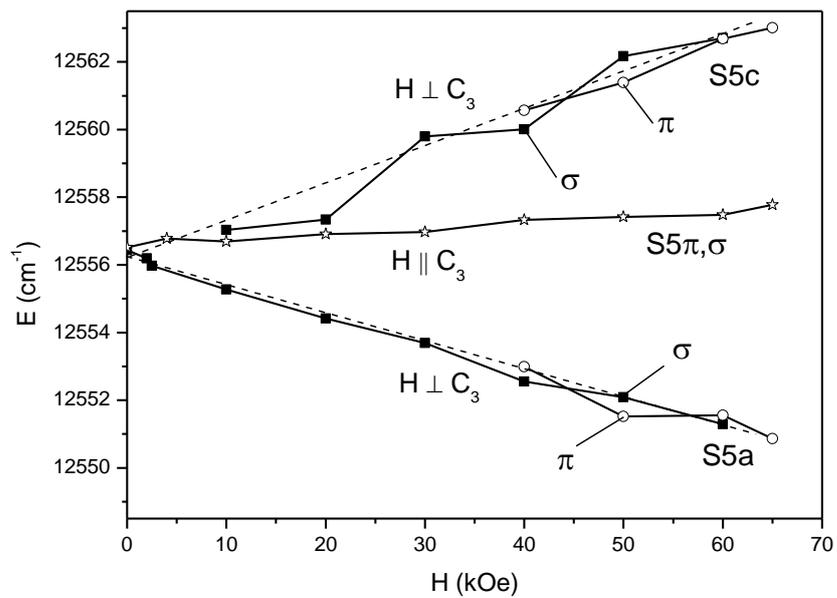

Fig. 25. Energies of the S5 transitions as a function of the magnetic field $H \perp C_3$ and $H \| C_3$.

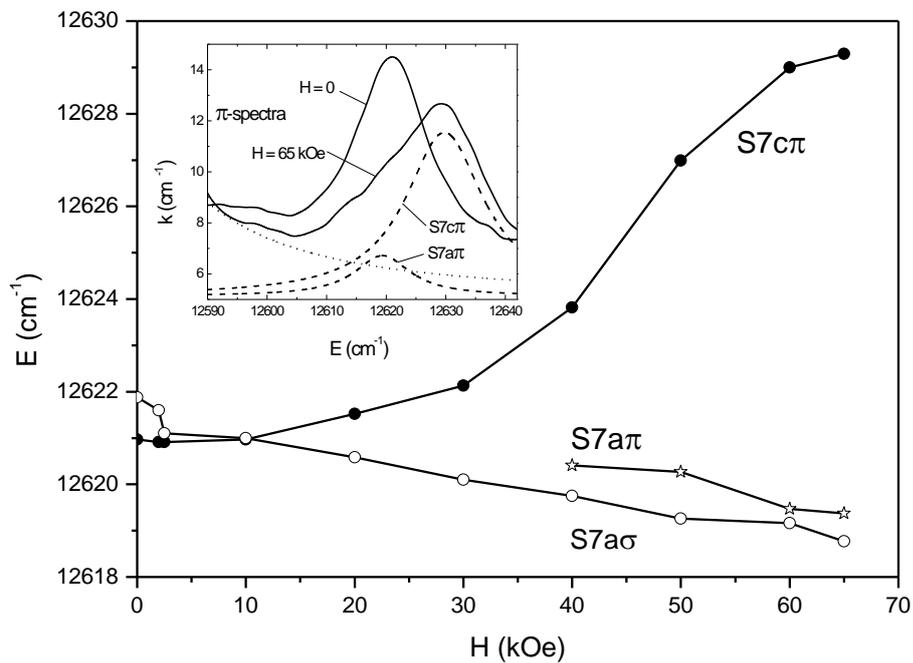

Fig. 26. Energies of the S7 transitions as a function of the magnetic field $H \perp C_3$. Inset: absorption spectra of the S7π transition in two magnetic fields $H \perp C_3$.

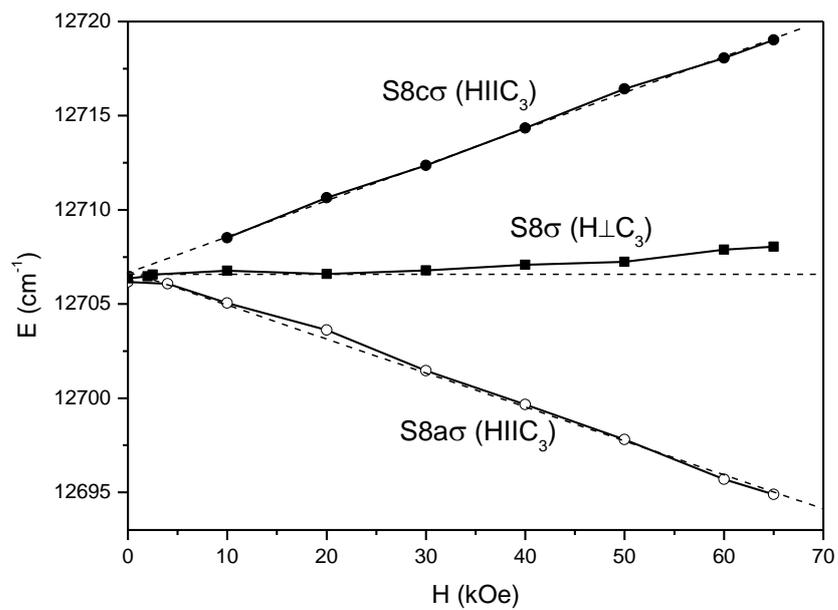

Fig. 27. Energies of the S8 transitions as a function of the magnetic field $H \perp C_3$ and $H \| C_3$.

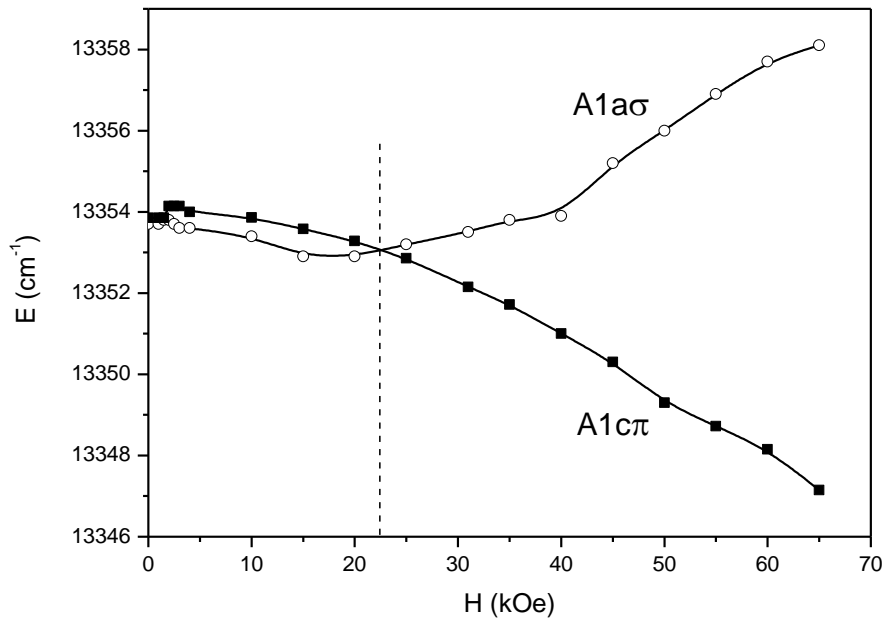

Fig. 28. Energies of the A1 transitions as a function of the magnetic field $H \perp C_3$.

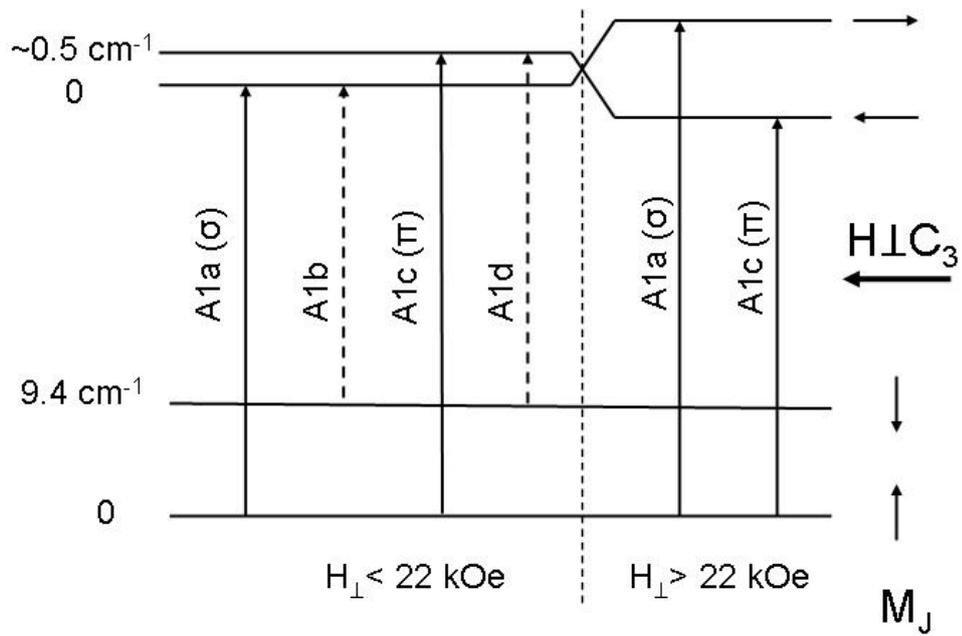

Fig. 29. Diagram of the A1 transitions.

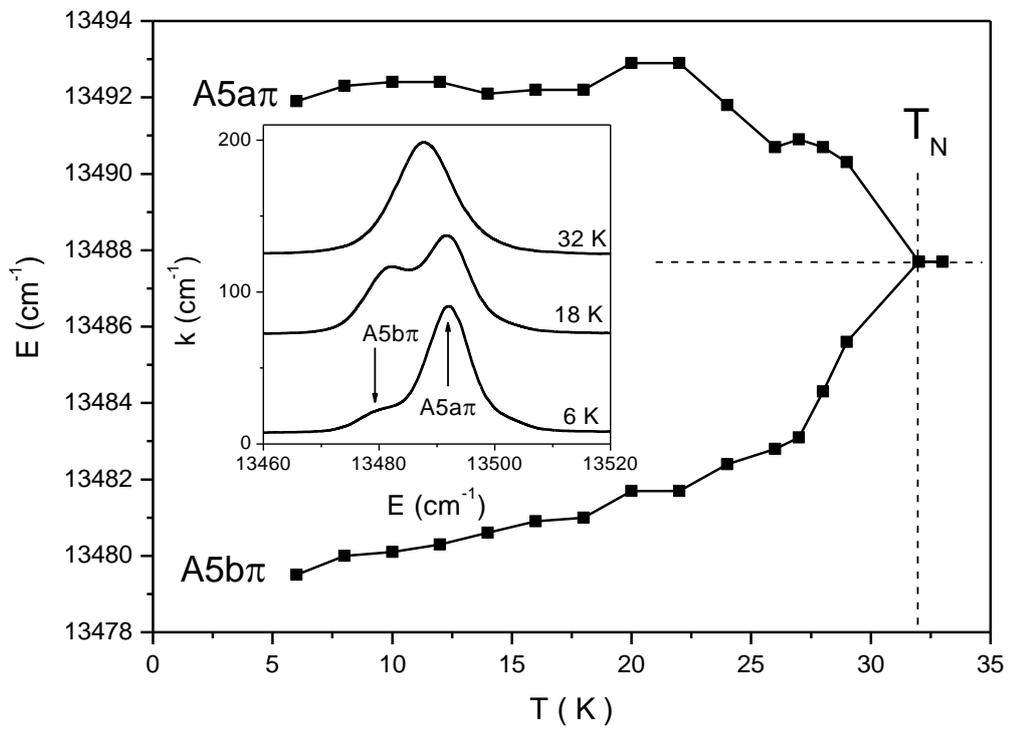

Fig. 30. Temperature dependence of the exchange splitting of the A5π transition.

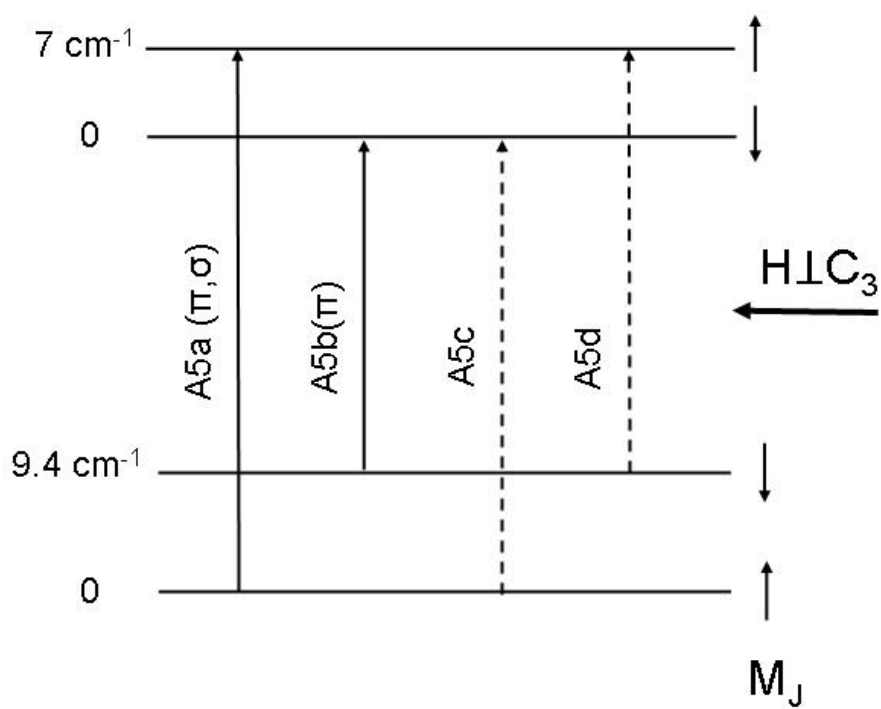

Fig. 31. Diagram of the A5 transitions.

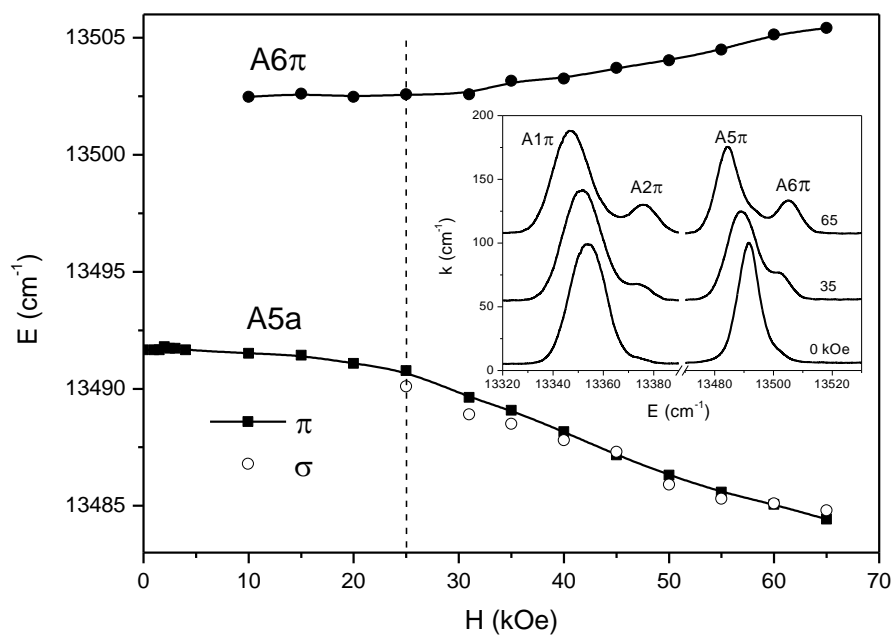

Fig. 32. Energies of the A5 and A6 transitions as a function of the magnetic field $H\perp C_3$. Inset: fragment of the A band spectrum at several magnetic fields $H\perp C_3$.

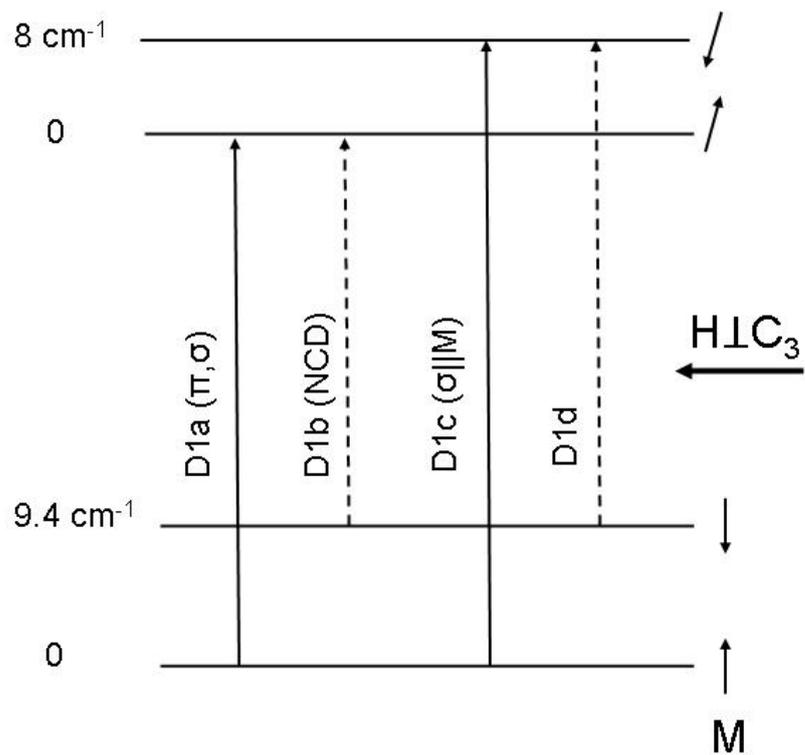

Fig. 33. Diagram of the D1 transitions.

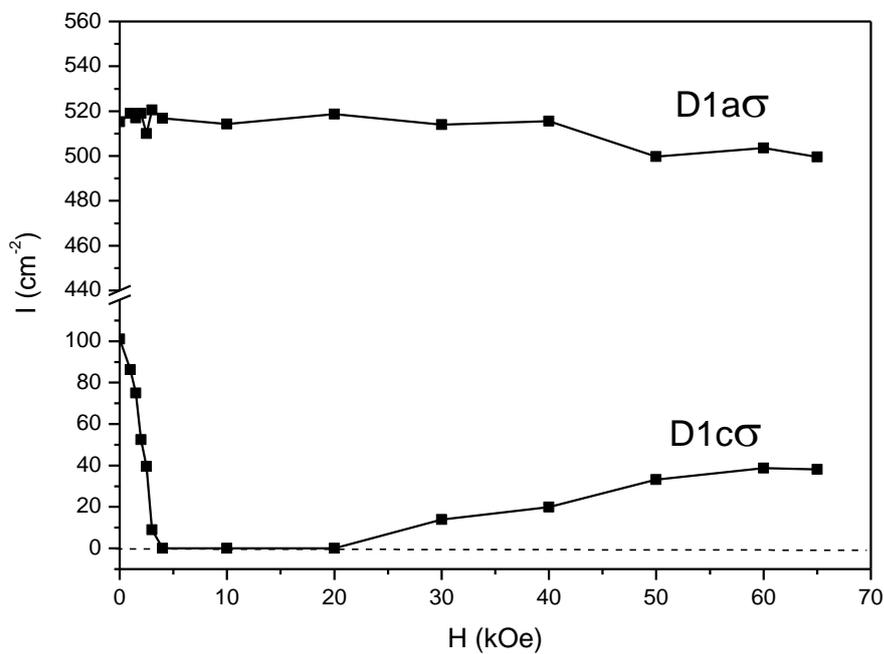

Fig. 34. Intensities of the D1a$\sigma$ and D1c$\sigma$ lines as a function of the magnetic field $H\perp C_3$.

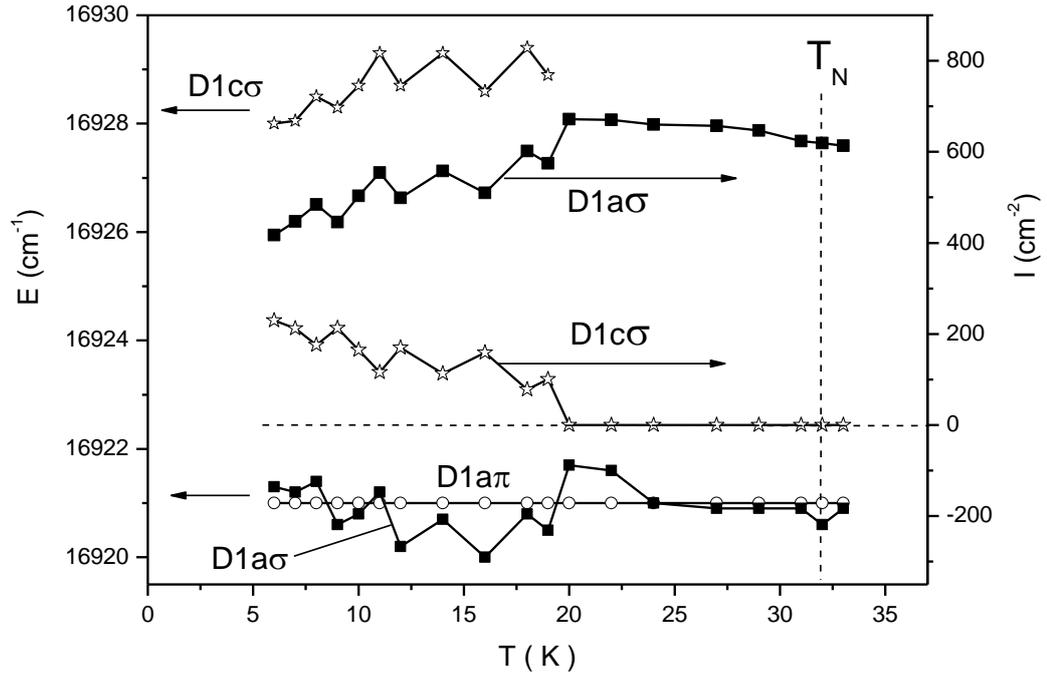

Fig. 35. Energies (*E*) and intensities (*I*) of the D1 lines as a function of temperature.

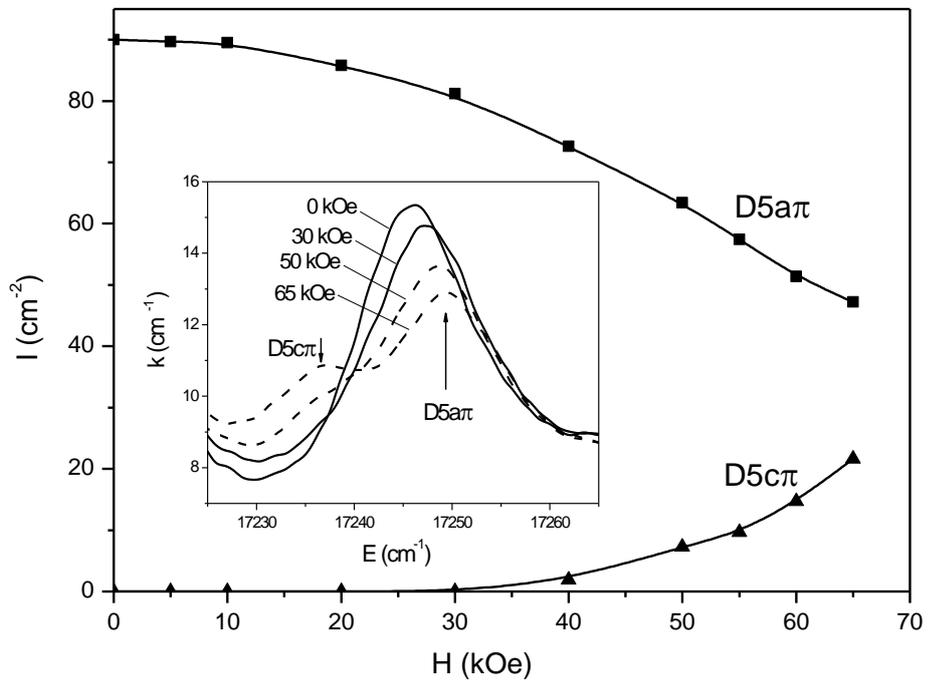

Fig. 36. Intensities of the D5aπ and D5cπ lines as a function of the magnetic field $H\|C_3$. Inset: transformation of the D5π transition spectrum in the magnetic field $H\|C_3$.

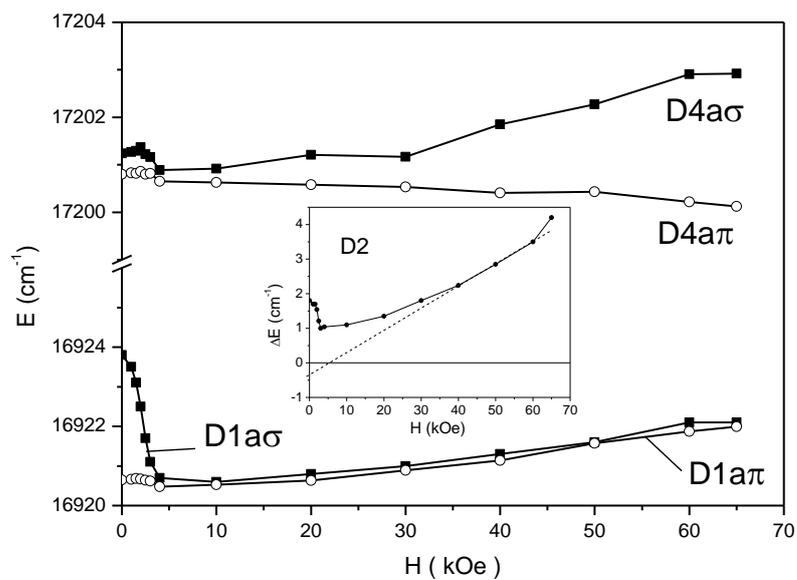

Fig. 37. Energies of the D1a and D4a transitions in two polarizations as a function of the magnetic field $H\perp C_3$. Inset: Splitting of the D2 line in the magnetic field $H\perp C_3$.

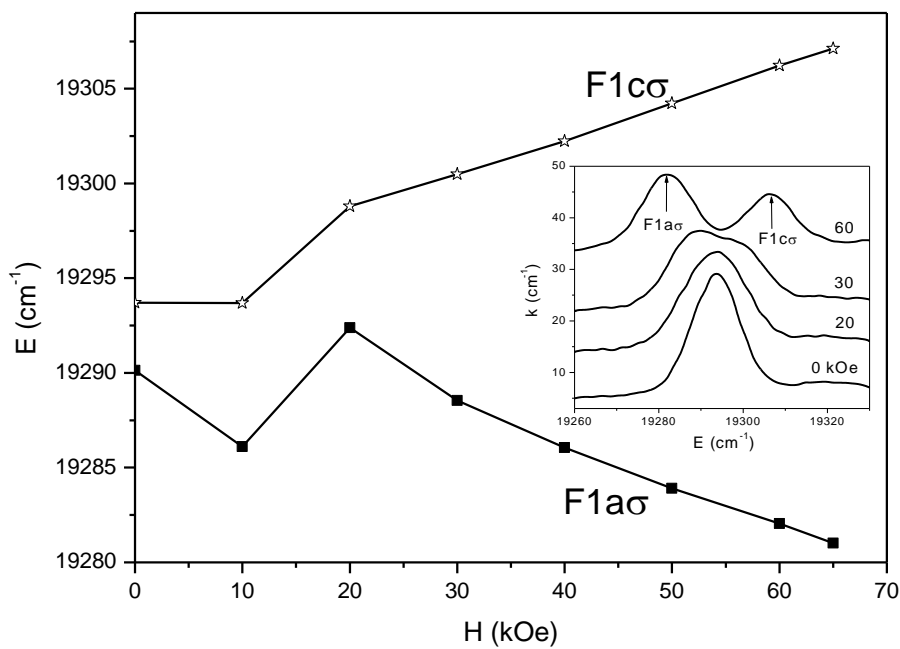

Fig. 38. Energies of the F1a$\sigma$ and F1c$\sigma$ transitions as a function of the magnetic field $H||C_3$. Inset: transformation of the F1$\sigma$ spectrum in the magnetic field $H||C_3$.

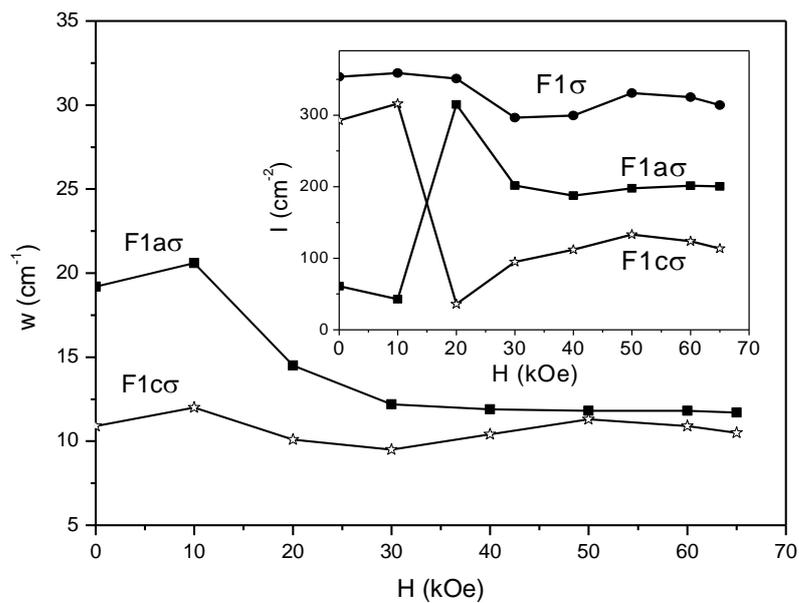

Fig. 39. Line widths and intensities (inset) of F1aσ and F1cσ transitions as a function of the magnetic field $H\|C_3$.

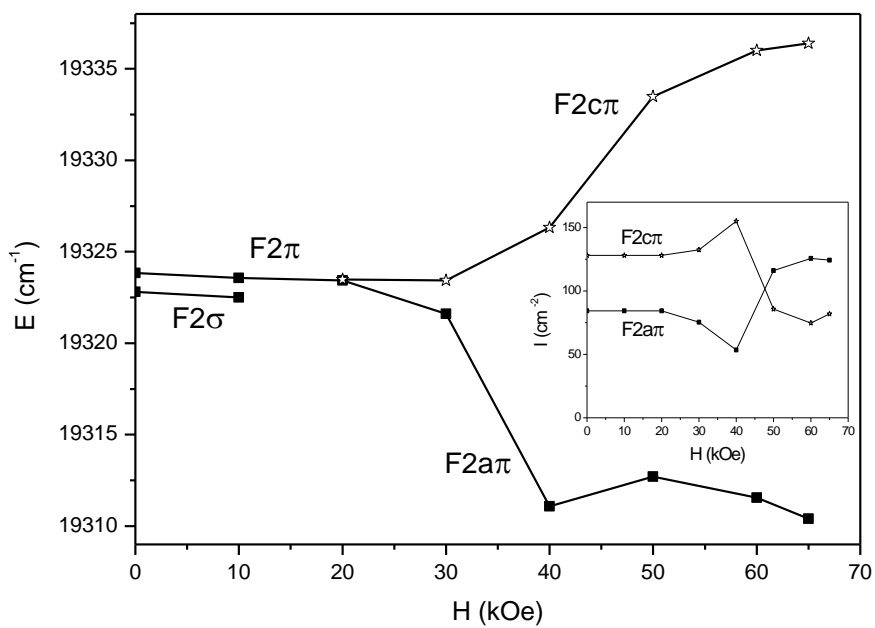

Fig. 40. Energies and intensities (inset) of the F2aπ and F2cπ transitions as a function of the magnetic field $H\|C_3$.

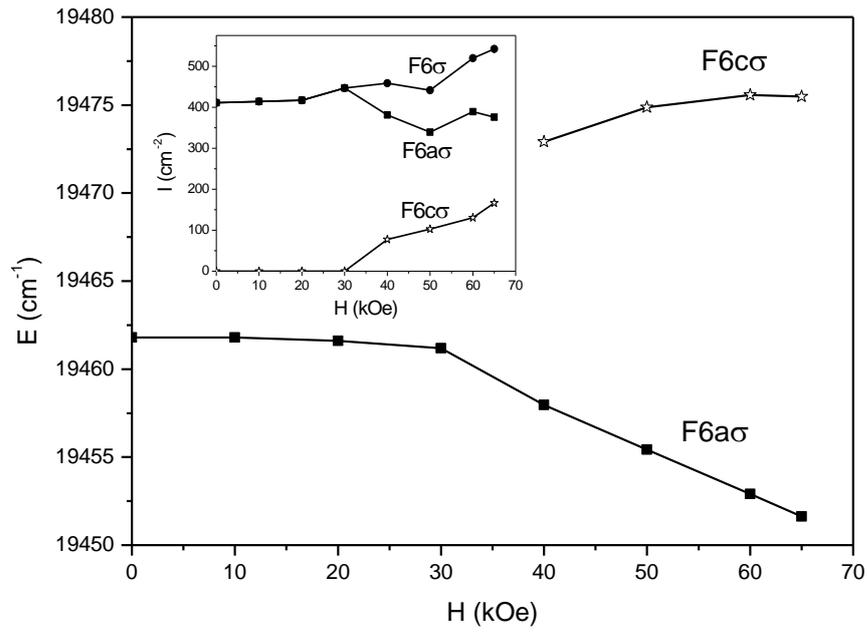

Fig. 41. Energies and intensities (inset) of the F6aσ and F6cσ transitions as a function of the magnetic field $H \| C_3$.